\documentclass{pasj01}
\usepackage{lscape}
\usepackage{color}
\usepackage{comment}
\usepackage{natbib, aas_macros}
\Received{$\langle$reception date$\rangle$}
\Accepted{$\langle$acception date$\rangle$}
\Published{$\langle$publication date$\rangle$}
\newcommand{\simgt}{\lower.5ex\hbox{$\; \buildrel > \over \sim \;$}}
\newcommand{\simlt}{\lower.5ex\hbox{$\; \buildrel < \over \sim \;$}}

\def\h70Msol{\mathrel{h_{70}^{-1}M_\odot}}
\begin{document}
\title{X-ray properties of high-richness CAMIRA clusters in the Hyper Suprime-Cam Subaru Strategic Program field}
 \author{Naomi \textsc{Ota}\altaffilmark{1}}
 \author{Ikuyuki \textsc{Mitsuishi}\altaffilmark{2}}
 \author{Yasunori \textsc{Babazaki}\altaffilmark{2}}
 \author{Hiroki \textsc{Akamatsu}\altaffilmark{3}}
 \author{Yuto \textsc{Ichinohe}\altaffilmark{4}}
 \author{Shutaro \textsc{Ueda}\altaffilmark{5}}
 \author{Nobuhiro \textsc{Okabe}\altaffilmark{6,7,8}}
 \author{Masamune \textsc{Oguri}\altaffilmark{9,10,11}}
 \author{Ryuichi \textsc{Fujimoto}\altaffilmark{14}}
 \author{Takashi \textsc{Hamana}\altaffilmark{12}}
 \author{Keita \textsc{Miyaoka}\altaffilmark{6}}
 \author{Satoshi \textsc{Miyazaki}\altaffilmark{12,13}}
 \author{Hanae \textsc{Otani}\altaffilmark{1}}
 \author{Keigo \textsc{Tanaka}\altaffilmark{14}}
 \author{Ayumi \textsc{Tsuji}\altaffilmark{14}}
 \author{Atsushi \textsc{Yoshida}\altaffilmark{2}}

\altaffiltext{1}{Department of Physics, Nara Women's University, Kitauoyanishi-machi, Nara, Nara 630-8506, Japan}
\altaffiltext{2}{Department of Physics, Nagoya University, Aichi 464-8602, Japan}
\altaffiltext{3}{SRON Netherlands Institute for Space Research, Sorbonnelaan 2, 3584 CA Utrecht, The Netherlands}
\altaffiltext{4}{Department of Physics, Rikkyo University, 3-34-1 Nishi-Ikebukuro, Toshima, Tokyo 171-8501, Japan}
%\altaffiltext{5}{Institute of Space and Astronautical Science (ISAS), JAXA, 3-1-1 Yoshinodai, Chuo, Sagamihara, Kanagawa 252-5210, Japan}
\altaffiltext{5}{Academia Sinica Institute of Astronomy and Astrophysics (ASIAA), No. 1, Section 4, Roosevelt Road, Taipei 10617, Taiwan}
\altaffiltext{6}{Department of Physical Science, Hiroshima University, 1-3-1 Kagamiyama, Higashi-Hiroshima, Hiroshima 739-8526, Japan}
\altaffiltext{7}{Hiroshima Astrophysical Science Center, Hiroshima University, 1-3-1, Kagamiyama, Higashi-Hiroshima, Hiroshima 739-8526, Japan}
\altaffiltext{8}{Core Research for Energetic Universe, Hiroshima University, 1-3-1, Kagamiyama, Higashi-Hiroshima, Hiroshima 739-8526, Japan}
\altaffiltext{9}{Research Center for the Early Universe, University of Tokyo, Tokyo 113-0033, Japan}
\altaffiltext{10}{Department of Physics, University of Tokyo, Tokyo 113-0033, Japan}
\altaffiltext{11}{Kavli Institute for the Physics and Mathematics of the Universe (Kavli IPMU, WPI), University
of Tokyo, Chiba 277-8582, Japan}
\altaffiltext{12}{National Astronomical Observatory of Japan, Mitaka, Tokyo 181-8588, Japan}
\altaffiltext{13}{Department of Astronomy, School of Science, Graduate University for Advanced Studies,
Mitaka, Tokyo 181-8588, Japan}
\altaffiltext{14}{Graduate School of Natural Science \& Technology, Kanazawa University, Kakuma-machi, Kanazawa, Ishikawa 920-1192, Japan}
\email{naomi@cc.nara-wu.ac.jp}

\KeyWords{cosmology: observations --- galaxies: clusters: intergalactic medium --- X-rays: galaxies: clusters}

\maketitle

\begin{abstract}
  We present the first results of a pilot X-ray study of 37 rich
  galaxy clusters at $0.1<z<1.1$ in the Hyper Suprime-Cam Subaru
  Strategic Program (HSC-SSP) field.  Diffuse X-ray emissions from
  these clusters were serendipitously detected in the XMM-Newton
  fields of view.  We systematically analyze X-ray images of 37
  clusters and emission spectra of a subsample of 17 clusters with
  high photon statistics by using the XMM-Newton archive data. The
  frequency distribution of the offset between the X-ray centroid or
  peak and the position of the brightest cluster galaxy was derived
  for the optical cluster sample. The fraction of relaxed clusters
  estimated from the X-ray peak offsets in 17 clusters is
  $29\pm11(\pm13)$\%, which is smaller than that of the X-ray cluster
  samples such as HIFLUGCS. Since the optical cluster search is immune
  to the physical state of X-ray-emitting gas, it is likely to cover a
  larger range of the cluster morphology. We also derived the
  luminosity-temperature relation and found that the slope is
  marginally shallower than those of X-ray-selected samples and
  consistent with the self-similar model prediction of 2. Accordingly,
  our results show that the X-ray properties of the optical clusters
  are marginally different from those observed in the X-ray
  samples. The implication of the results and future prospects are
  briefly discussed.
\end{abstract}

\section{Introduction}
The Hyper Suprime-Cam Subaru Strategic Program
\citep[HSC-SSP;][]{HSC1styr,HSC1styrOverview,HSCPhotoz18,Bosch18} is
an ongoing wide-field imaging survey that uses the HSC
\citep{Miyazaki12,Miyazaki15,Miyazaki18,Komiyama18,Kawanomoto18,Furusawa18}
mounted on the prime focus of the Subaru Telescope.  The HSC-SSP
survey has three different layers, Wide, Deep, and Ultra-deep.  The
wide layer takes five-band ($grizy$) and deep ($r\simlt26$~AB~mag)
imaging over $1400$~deg$^2$.  To date, the survey covers $456$~deg$^2$
with non-full-depth and $178$~deg$^2$ with the full-depth and
full-color \citep{HSC1styr}.

The deep and multi-band HSC-SSP imaging gives us a unique opportunity
to conduct a systematic search of optical galaxy clusters. In fact,
\cite{Oguri18} discovered 2000 galaxy clusters with richness
$\hat{N}_{\rm mem}>15$ in $\sim232$~deg$^2$, by applying the CAMIRA
algorithm developed by \cite{Oguri14b}.  The galaxy clusters are
discovered as concentrations of red-sequence galaxies by applying a
compensated spatial filter to the three-dimensional richness map.  The
accuracy of photometric redshifts of the CAMIRA clusters is $\Delta
z/(1+z)\sim0.01$.

The CAMIRA catalog features a wide redshift coverage and a low mass
limit, which therefore provides us with an unprecedented cluster
sample including high-redshift objects.  Because the limiting
magnitudes of the HSC-SSP survey is much deeper than those of the
Sloan Digital Sky Survey (SDSS) and Dark Energy Survey (DES), the
galaxy clusters can be securely identified up to $z\sim 1.1$, in
contrast with the SDSS \citep[$z\sim0.4$;][]{Oguri14b,Rykoff14} and
the DES \citep[$z\sim0.8$;][]{Rykoff16}. The redshift range is
comparable to those covered by Sunyaev-Zel'dovich effect (SZE)
surveys, which used the South Pole Telescope \citep[][]{SPTSZ15} and
the Atacama Cosmology Telescope \citep{Hilton17}. The richness
$\hat{N}_{\rm mem}\sim 15$ roughly corresponds to $M_{200m}\sim
10^{14}h^{-1}M_\odot$ \citep{Oguri18} and is equivalent to
$M_{500}\sim7\times 10^{13}\h70Msol$ if we assume a median halo
concentration of $c_{200\rm m}=6$ \citep{Diemer15}.  The detection
limit of the cluster mass for the CAMIRA clusters is then much lower
than those of the SZE clusters \citep[$M_{500}\sim3.5\times
10^{14}\h70Msol$;][]{SPTSZ15}.

To understand the gas physics and establish scaling relations between
cluster mass and X-ray observables in preparation for future
cosmological research, it is important to systematically study the
X-ray properties of the optically-selected clusters and compare them
with other multi-wavelength surveys.  To date, a number of systematic
cluster observations \citep[see,
e.g.,][]{Vikhlinin06,Zhang08,Sun09,Martino14,Mahdavi13,
  Donahue14,vonderLinden14,Okabe14b,Hoekstra15,Smith16,Mantz16} have
been conducted by referring to cluster catalogs constructed from the
ROSAT All Sky Survey \citep[RASS; e.g.,][]{Bohringer01}.  More
recently, statistical studies use the cutting-edge X-ray surveys
\citep[e.g.,][]{XXL16}, SZE \citep[e.g.,][]{Sanders18,Bulbul19}, or
optical techniques \citep[e.g.,][]{Hicks08,Hicks13,Takey13}.  Since
different survey techniques have their own selection functions, some
systematic differences may appear in their observed cluster properties
and scaling relations. If this happens, a selection bias issue arises,
which eventually leads to a difficulty in constraining the
cosmological models using the cluster mass function \citep[see,
e.g.,][]{Allen11,Giodini13}. This will have an impact on
interpretation of the upcoming eROSITA \citep{Merloni12} and other
ongoing/future large-scale cluster surveys.

A useful measure of the cluster dynamical state is given by the offset
between the location of the brightest cluster galaxy (BCG) and the
X-ray centroid or X-ray peak \citep[e.g.,][]{Katayama03}.  The X-ray
centroid (or peak) offset is sometimes used to classify the clusters
into relaxed and disturbed clusters
\citep{Mann12,Mahdavi13,Rossetti16}. \cite{Rossetti16} showed that the
fraction of relaxed clusters is smaller in the Planck sample than that
in the X-ray samples, indicating that SZE and X-rays surveys of galaxy
clusters are affected by the different selection effects.  In this
way, the X-ray centroid offset is useful not only to characterize the
cluster dynamical state but also to study the selection effect. While
the offsets between optical and X-ray centers have been used to study
the misidentification of central galaxies in optical cluster finding
algorithms \citep[e.g.,][]{Rozo14,Rykoff16,Oguri18}, dynamical states
of optically selected clusters based on offset distributions have not
yet been fully explored.  To address this situation, this paper
presents a systematic measurement of the centroid offset in the
optical sample.

We thus carried out a systematic X-ray analysis of the CAMIRA clusters
with high optical richness using the XMM-Newton archival data.
Section~\ref{sec:sample} presents the sample selection and
section~\ref{sec:analysis} describes the data analyses regarding
centroid determination and spectral analysis.
Section~\ref{sec:results} derives the centroid offset and the
luminosity-temperature relation, and section~\ref{sec:discussion}
discusses the implication of the results. Finally
section~\ref{sec:summary} summarizes the results and briefly discusses
the future prospects of this X-ray follow-up project.

The cosmological parameters are $\Omega_{m0}=0.28$,
$\Omega_\Lambda=0.72$ and $h=0.7$ throughout this paper, and we use
the proto-solar abundance table from \cite{Lodders09}.  Unless
otherwise noted, the quoted errors represent the $1\sigma$ statistical
uncertainties.

\section{Sample}\label{sec:sample}
\begin{figure}[htb]
 \begin{center}
\includegraphics[width=7cm,angle=90]{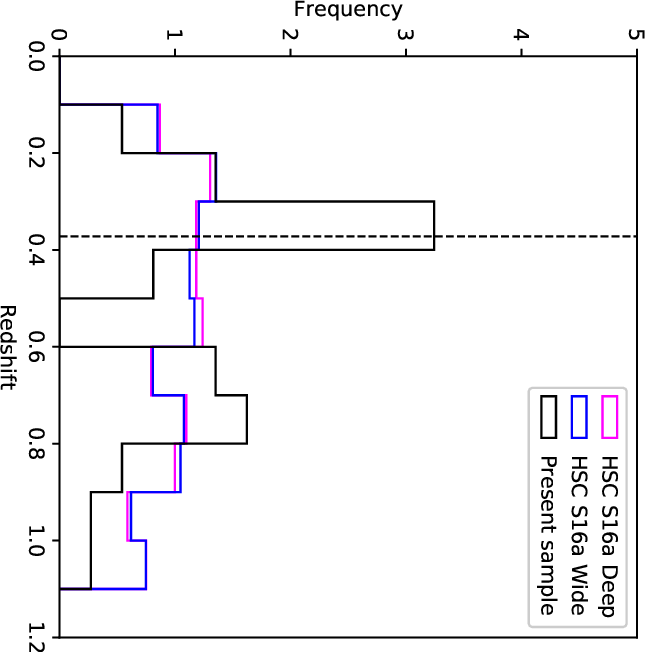}
 \end{center}
 \caption{Redshift distributions of the present sample (black), CAMIRA
   HSC S16A Wide (blue) and Deep (magenta) cluster catalogs
   \citep{Oguri18}. The binsize is $\Delta z=0.1$ and each histogram
   is normalized such that the integral over the range is unity. The
   vertical dashed line indicates the median redshift of the present
   sample, $\tilde{z}=0.37$. }\label{fig:sample}
\end{figure}

\begin{figure*}[htb]
 \begin{center}
  \includegraphics[width=8cm]{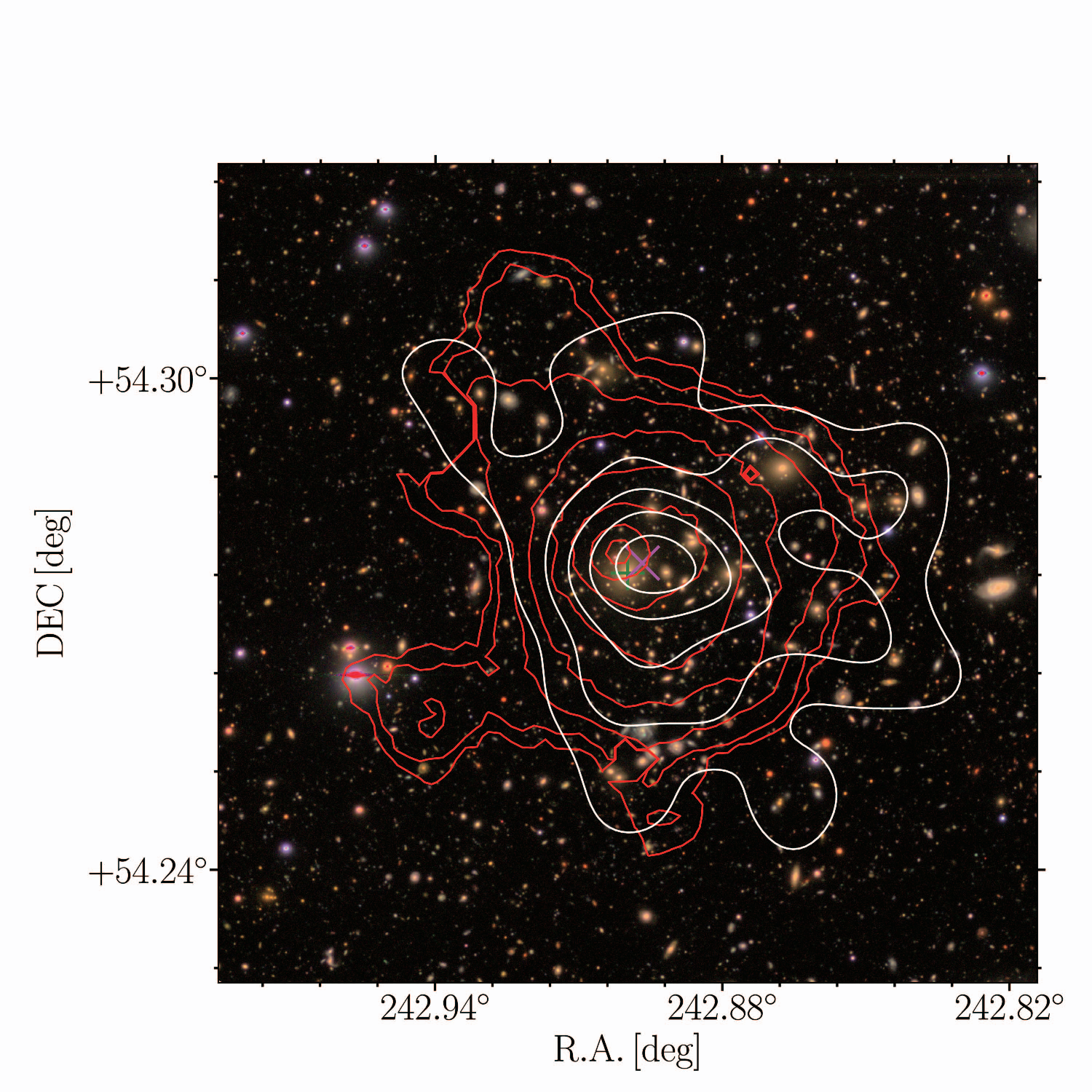}
 \includegraphics[width=8cm]{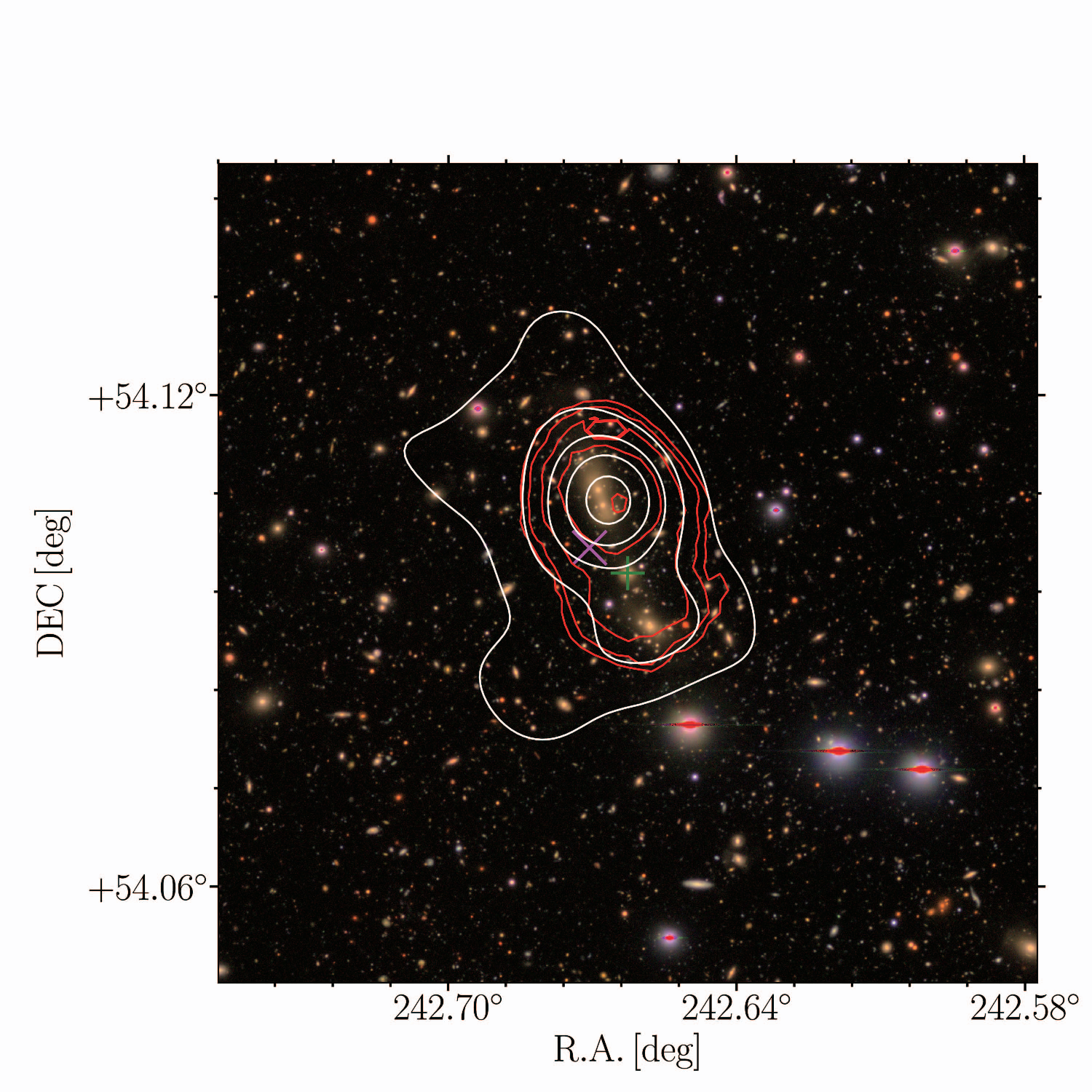}
 \end{center}
 \caption{Examples of HSC I-band images of the CAMIRA clusters,
   HSC~J161136+541635 at $z=0.332$ (left panel) and HSC~J161039+540554
   at $z=0.330$ (right panel). In each panel, the X-ray centroid and
   BCG positions are marked with a magenta ``$\times$'' and green
   ``+'', respectively. The white contours are linearly spaced by half
   of the average height of galaxy density maps over all CAMIRA
   clusters at the same redshift. The red contours for X-ray emission
   are ten levels logarithmically spaced from $[10-1000]\,{\rm
     cts\,s^{-1}\,deg^{-2}}$.  }\label{fig:image}
\end{figure*}

The CAMIRA catalog comprises 2086 clusters at $0.1<z<1.1$ in the S16A
Wide and Deep fields \citep{Oguri18}.  We cross-correlated the CAMIRA
catalog with the 3XMM-DR7 catalog \citep{Rosen16} to find that there
are $>300$ X-ray sources within $60\arcsec$ from the optical centers.
We then excluded a $\sim25\,{\rm deg}^2$ XXL survey region overlapped
with that of the HSC-SSP survey from the above search result; an X-ray
study in the XXL field is to be done through the HSC-XXL external
collaboration. To do the systematic X-ray analysis of high-richness
clusters, we construct the sample by selecting objects with richness
$\hat{N}_{\rm mem}>20$.  This richness range corresponds to the
cluster mass $M_{500}\simgt7\times10^{13}~{\rm M_{\odot}}$
\citep{Okabe18}.  Therefore, as listed in Table~\ref{tbl:sample}, the
present sample consists of 37 clusters at $0.14 < z < 1.09$, whose
distribution is overlaid on that of the CAMIRA clusters with
$\hat{N}_{\rm mem}>20$ (Figure~\ref{fig:sample}).  Except for
HSC~J141508-002936 at $z=0.14$ (alternative name is Abell~1882) for
which X-ray data were taken by pointed XMM-Newton observations
\citep{Miyaoka18}, X-ray emissions from these clusters are
serendipitously detected inside the XMM-Newton fields of view. The
average (median) redshift is 0.50 (0.37).  Examples of HSC images of
the CAMIRA clusters are shown in Figure~\ref{fig:image}.

Because we require typically more than 1000 cluster-photon counts so
as to enable X-ray spectroscopic measurements of the gas temperature
and luminosity, the sample is subdivided into two: 17 clusters with
$>1000$~cluster-photon counts and 20 clusters with
$<1000$~counts\footnote{The sum of X-ray counts observed by XMM-Newton
  EPIC sensors.}, which are listed in the 1--17th and 18--37th rows of
Table~\ref{tbl:sample}, respectively. The former (latter) subsample
has the average redshift of 0.33 (0.67). The total sample covers the
equivalent redshift range of the CAMIRA catalog, but has a lower
average redshift.  The Kolmogorov-Smirnov (K-S) test gave the
probability that the two samples are from the same redshift
distribution as $p=0.13$ (the K-S parameter $D=0.19$), while the K-S
test yielded $p=0.002$ ($D=0.31$) for the richness distribution.  At
the 5\% significance level, the null hypothesis is not rejected
(rejected) in the former (latter) case. We found that the K-S
statistics has been improved by including the twenty clusters with low
counts, however, the above test suggests that the present measurement
is subject to a selection bias. Ideally, a sample of purely
optically-selected clusters will be constructed by cross-matching the
CAMIRA catalog with the XXL \citep{Pierre16} or future eROSITA cluster
catalog, which enables unbiased study of X-ray properties of the
optically-selected sample including objects with upper limits where no
X-ray emission is detected.  This is not possible for the present
study, and the selection bias will be examined in
section~\ref{subsec:lt}.

Table~\ref{tbl:sample} lists the location of BCGs identified by the
CAMIRA algorithm \citep{Oguri18}. Note that for 4 out of 37 clusters,
the BCGs are clearly misidentified by the CAMIRA algorithm for either
one of the following reasons; i) there is another elliptical galaxy
near the cluster center brighter than that listed in the CAMIRA
catalog, ii) the BCG in the CAMIRA catalog lies outside the X-ray core
and there is equivalently bright one inside the core.  Since we are
interested in physical offsets between BCGs and X-ray peaks rather
than miscentering of optical cluster finding algorithms, we correct
the BCG coordinates for these four (HSC~J140309-001833,
HSC~J021427-062720, HSC~J100049+013820, HSC~J222210-004421) by visual
inspection of their HSC images.

\begin{table*}[hbt]
  \tbl{Sample list.}{%
  \begin{tabular}{llllllllll}\hline\hline
Cluster  & $z$ & $\hat{N}_{\rm mem}$$^{\mathrm{a}}$ & $R_{500}$&  BCG position & X-ray centroid & $D_{\rm XC}$$^{\mathrm{b}}$ & $D_{\rm XP}$$^{\mathrm{c}}$  & OBSID$^{\mathrm{d}}$  & Exposure$^{\mathrm{e}}$  \\ 
	    & & & (Mpc/\arcsec) & RA, Dec (deg)  & RA, Dec (deg) & (kpc) & (kpc) & & M1, M2, PN \\ \hline
HSC~J142624-012657	&	0.460	&	69.7	&	0.835	/	142	&	216.6011	,	-1.4492	&	216.5989	,	-1.4502	&	52	&	23 (23,60)		&	0674480701	&	12.3	,	0.6	,	9.1	\\
HSC~J021115-034319	&	0.745	&	52.3	&	0.653	/	88	&	32.8135	 ,	-3.7219	&	32.8132	,	-3.7225	&	18	&	60 (60,77)		&	0655343861	&	7.7	,	13.0	,	3.4	\\
HSC~J095939+023044	&	0.730	&	51.7	&	0.657	/	90	&	149.9132	,	2.5122	&	149.9188	,	2.5193	&	239	&	196 (190,287)	&	0203361701	&	30.1	,	30.2	,	24.3	\\
HSC~J161136+541635	&	0.332	&	48.3	&	0.807	/	168	&	242.8998	,	54.2763	&	242.8981	,	54.2771	&	22	&	38 (38,43)		&	0059752301	&	4.9	,	4.7	,	2.9	\\
HSC~J090914-001220	&	0.303	&	46.5	&	0.811	/	180	&	137.3075	,	-0.2056	&	137.3086	,	-0.2052	&	20	&	23 (22,40)		&	0725310142	&	2.5	,	2.7	,	2.4	\\
HSC~J141508-002936	&	0.144	&	43.0	&	0.860	/	340	&	213.7850	,	-0.4932	&	213.7835	,	-0.4891	&	40	&	50 (39,62)		&	0145480101	&	11.0	,	11.8	,	6.9	\\
HSC~J140309-001833	&	0.449	&	39.7	&	0.715	/	124	&	210.7876	,	-0.3091	&	210.7939	,	-0.3069	&	76	&	36 (15,36)		&	0606430501	&	20.4	,	21.1	,	13.3	\\
HSC~J095737+023426	&	0.372	&	37.4	&	0.734	/	142	&	149.4043	,	2.5738	&	149.4050	,	2.5748	&	23	&	14 (14,44)		&	0203362201	&	28.9	,	29.0	,	12.3	\\
HSC~J022135-062618	&	0.300	&	35.7	&	0.754	/	169	&	35.3947	,	-6.4384	&	35.4069	,	-6.4457	&	228	&	10 (10,13)		&	0655343837	&	2.6	,	2.6	,	2.2	\\
HSC~J232924-004855	&	0.310	&	35.2	&	0.746	/	163	&	352.3487	,	-0.8154	&	352.3495	,	-0.8147	&	17	&	44 (44,44)		&	0673002346	&	3.5	,	3.8	,	1.8	\\
HSC~J022512-062259	&	0.202	&	33.0	&	0.775	/	232	&	36.3012	,	-6.3831	&	36.2985	,	-6.3811	&	40	&	246 (246,262)	&	0655343836	&	2.6	,	2.5	,	2.2	\\
HSC~J021427-062720	&	0.246	&	31.3	&	0.746	/	192	&	33.6071	,	-6.4607	&	33.6186	,	-6.4562	&	171	&	15 (13,34)		&	0655343859	&	2.5	,	2.7	,	2.0	\\
HSC~J161039+540554	&	0.330	&	29.5	&	0.702	/	147	&	242.6626	,	54.0983	&	242.6697	,	54.1031	&	109	&	144 (144,162)	&	0059752301	&	4.9	,	4.8	,	2.9	\\
HSC~J095903+025545	&	0.332	&	26.4	&	0.679	/	142	&	149.7614	,	2.9291	&	149.7620	,	2.9214	&	133	&	8 (8,32)		&	0203361601	&	19.1	,	0.0	,	8.6	\\
HSC~J100049+013820	&	0.228	&	23.2	&	0.692	/	189	&	150.1898	,	1.6573	&	150.1923	,	1.6592	&	41	&	94 (87,102)	&	0302351001	&	37.6	,	38.9	,	28.0	\\
HSC~J090743+013330	&	0.172	&	23.1	&	0.711	/	242	&	136.9295	,	1.5583	&	136.9540	,	1.5564	&	260	&	14 (14,14)		&	0725310156	&	2.6	,	2.7	,	2.4	\\
HSC~J095824+024916	&	0.341	&	20.1	&	0.625	/	128	&	149.6001	,	2.8212	&	149.6001	,	2.8214	&	5	&	9 (9,13)		&	0203362101	&	59.4	,	59.5	,	51.1	\\
	 \hline
HSC~J090754+005732	&	0.692	&	43.5	&	0.639	/	89	&	136.9765	,	0.9590	&	136.9769	,	0.9632	&	111	&	131 (131,167)	&	0725310159	&	2.7	,	2.7	,	2.2	\\
HSC~J090541+013226	&	0.666	&	39.1	&	0.629	/	89	&	136.4217	,	1.5406	&	136.4213	,	1.5378	&	74	&	52 (52,86)		&	0725310131	&	8.5	,	8.3	,	7.6	\\
HSC~J232924-004855	&	0.310	&	35.2	&	0.746	/	163	&	352.3487	,	-0.8154	&	352.3487	,	-0.8086	&	113	&	31 (10,31)		&	0673002345	&	2.1	,	3.4	,	0.6	\\
HSC~J222210-004421	&	0.956	&	32.6	&	0.505	/	63	&	335.5444	,	-0.7623	&	335.5386	,	-0.7592	&	188	&	289 (211,493)	&	0670020201	&	4.2	,	4.0	,	2.3	\\
HSC~J100221+032807	&	1.088	&	31.8	&	0.466	/	56	&	150.5864	,	3.4687	&	150.5873	,	3.4654	&	102	&	363 (311,363)	&	0743110701	&	0.0	,	29.7	,	24.3	\\
HSC~J221211-000821	&	0.350	&	30.5	&	0.701	/	141	&	333.0477	,	-0.1391	&	333.0477	,	-0.1381	&	17	&	75 (51,75)		&	0655346840	&	0.0	,	2.8	,	2.6	\\
HSC~J160424+430438	&	0.856	&	30.4	&	0.525	/	68	&	241.1001	,	43.0771	&	241.0993	,	43.0778	&	26	&	42 (42,42)		&	0025740401	&	12.2	,	12.4	,	6.1	\\
HSC~J100300+013152	&	0.676	&	30.3	&	0.582	/	82	&	150.7505	,	1.5310	&	150.7484	,	1.5340	&	93	&	80 (47,80)		&	0203360501	&	26.8	,	0.0	,	18.8	\\
HSC~J142203-000402	&	0.630	&	30.1	&	0.597	/	87	&	215.5124	,	-0.0672	&	215.5195	,	-0.0647	&	187	&	352 (339,352)	&	0651740801	&	0.0	,	7.2	,	4.0	\\
HSC~J090419+020641	&	0.783	&	29.9	&	0.545	/	72	&	136.0793	,	2.1114	&	136.0797	,	2.1125	&	30	&	29 (17,293)	&	0725310152	&	2.7	,	0.0	,	2.3	\\
HSC~J220625+013905	&	0.281	&	27.4	&	0.706	/	165	&	331.6036	,	1.6514	&	331.6068	,	1.6591	&	129	&	15 (15,17)		&	0655346835	&	2.8	,	3.0	,	0.9	\\
HSC~J221422+004706	&	0.308	&	26.5	&	0.689	/	151	&	333.5930	,	0.7850	&	333.5857	,	0.7879	&	129	&	37 (37,47)		&	0655346839	&	3.0	,	3.1	,	2.6	\\
HSC~J090806+011956	&	0.672	&	26.0	&	0.558	/	79	&	137.0261	,	1.3321	&	137.0190	,	1.3238	&	279	&	243 (205,243)	&	0725310157	&	0.8	,	0.0	,	3.9	\\ % mos
HSC~J090509+012428	&	0.704	&	26.0	&	0.548	/	76	&	136.2884	,	1.4079	&	136.2908	,	1.4083	&	64	&	129 (99,165)	&	0725310131	&	8.5	,	8.3	,	7.6	\\
HSC~J022246-061703	&	0.772	&	24.3	&	0.517	/	69	&	35.6923	,	-6.2842	&	35.6894	,	-6.2779	&	189	&	211 (211,211)	&	0655343837	&	2.6	,	2.6	,	2.2	\\
HSC~J222121-004630	&	0.337	&	23.3	&	0.654	/	135	&	335.3380	,	-0.7751	&	335.3446	,	-0.7715	&	131	&	167 (148,167)	&	0670020201	&	4.2	,	4.0	,	2.3	\\
HSC~J221726-001020	&	0.325	&	21.8	&	0.645	/	136	&	334.3594	,	-0.1724	&	334.3585	,	-0.1726	&	15	&	171 (171,193)	&	0673000144	&	4.2	,	3.9	,	3.6	\\
HSC~J090602+011443	&	0.790	&	21.3	&	0.493	/	65	&	136.5097	,	1.2452	&	136.5127	,	1.2518	&	196	&	183 (183,220)	&	0725310149	&	0.0	,	2.4	,	1.9	\\
HSC~J221538+004227	&	0.441	&	20.7	&	0.597	/	104	&	333.9099	,	0.7074	&	333.9049	,	0.7137	&	166	&	77 (77,91)		&	0673000135	&	4.1	,	4.1	,	3.7	\\
HSC~J141648+521039	&	0.809	&	20.3	&	0.480	/	63	&	214.2018	,	52.1776	&	214.2007	,	52.1805	&	82	&	263 (191,263)	&	0127921001	&	49.0	,	48.9	,	0.0	\\
\hline
 \end{tabular}}\label{tbl:sample}
 \begin{tabnote}
   The sample is subdivided to 17 clusters with $>1000$ cluster-photon
   counts (1--17 rows) and 20 clusters with $<1000$~counts (18--37
   rows). $^{\mathrm{a}}$ Richness. $^{\mathrm{b}}$Centroid offset
   (see section \ref{subsec:centroid} for
   definition). $^{\mathrm{c}}$Peak offset (see
   section~\ref{subsec:centroidoffset} for definition).  The error
   range estimated by changing the smoothing scale of X-ray image is
   indicated in the parenthesis (see text). $^{\mathrm{d}}$The
   XMM-Newton observation id. $^{\mathrm{e}}$ The XMM-Newton
   EPIC-MOS1(M1), MOS2(M2), and PN exposure time after data filtering
   (ksec).
 \end{tabnote}
\end{table*}

\section{Analysis}\label{sec:analysis}
\subsection{Data reduction}\label{subsec:reduction}
Observation data files were retrieved from the XMM-Newton Science
Archive\footnote{http://nxsa.esac.esa.int} and reprocessed with the
XMM-Newton Science Analysis System v15.0.0 and the Current Calibration
Files. The data reduction, including flare screening, point source
detection, and estimation of the quiescent particle background, was
done in the standard manner by using the XMM Extended Source Analysis
Software [ESAS; \cite{Snowden08}; see also \cite{Miyaoka18}]. In the
following analysis, the detected point sources were excluded from the
EPIC data.

\subsection{Centroid determination}\label{subsec:centroid}
The X-ray centroid of each cluster was determined from the mean of the
photon distribution in an aperture circle of radius $R_{500}$. This
analysis used the 0.4--2.3~keV EPIC composite image (one image pixel
is 5\arcsec). Here, $R_{500}$ was calculated by substituting
$\hat{N}_{\rm mem}$ in Table~\ref{tbl:sample} in the $R-\hat{N}_{\rm
  mem}$ relation, which was deduced from the $R-T$ relation
\citep{Arnaud05} and the $T-\hat{N}_{\rm mem}$ relation
\citep{Oguri18}. Starting with the optical center, we iterated the
centroid search until its position converged within 5\arcsec.  If
contaminating point sources remained in the circle, we excluded the
region centered at the sources and their symmetric positions with
respect to the centroid determined by the previous iteration so as not
to affect the centroid determination \citep{Ota04}.  The result is
listed in Table~\ref{tbl:sample}. The offset between X-ray centroid
and BCG position is presented in section~\ref{subsec:centroidoffset}.

\subsection{Spectral analysis}\label{subsec:spec}
To evaluate the gas temperature and bolometric luminosity, we derive
the X-ray spectra by extracting the EPIC data from a circular region
within a radius of $R_{500}$ centered on the X-ray centroid.  For 17
clusters with sufficient photon statistics, the spectra were rebinned
so that each spectral bin contains over 25 counts. After subtracting
the quiescent particle background, the observed spectra of the EPIC
MOS/PN cameras in the 0.3--10/0.4--10~keV band were simultaneously fit
by using XSPEC 12.9.1 \citep{Arnaud96}.

\begin{figure}[htb]
 \begin{center}
  \includegraphics[width=7cm]{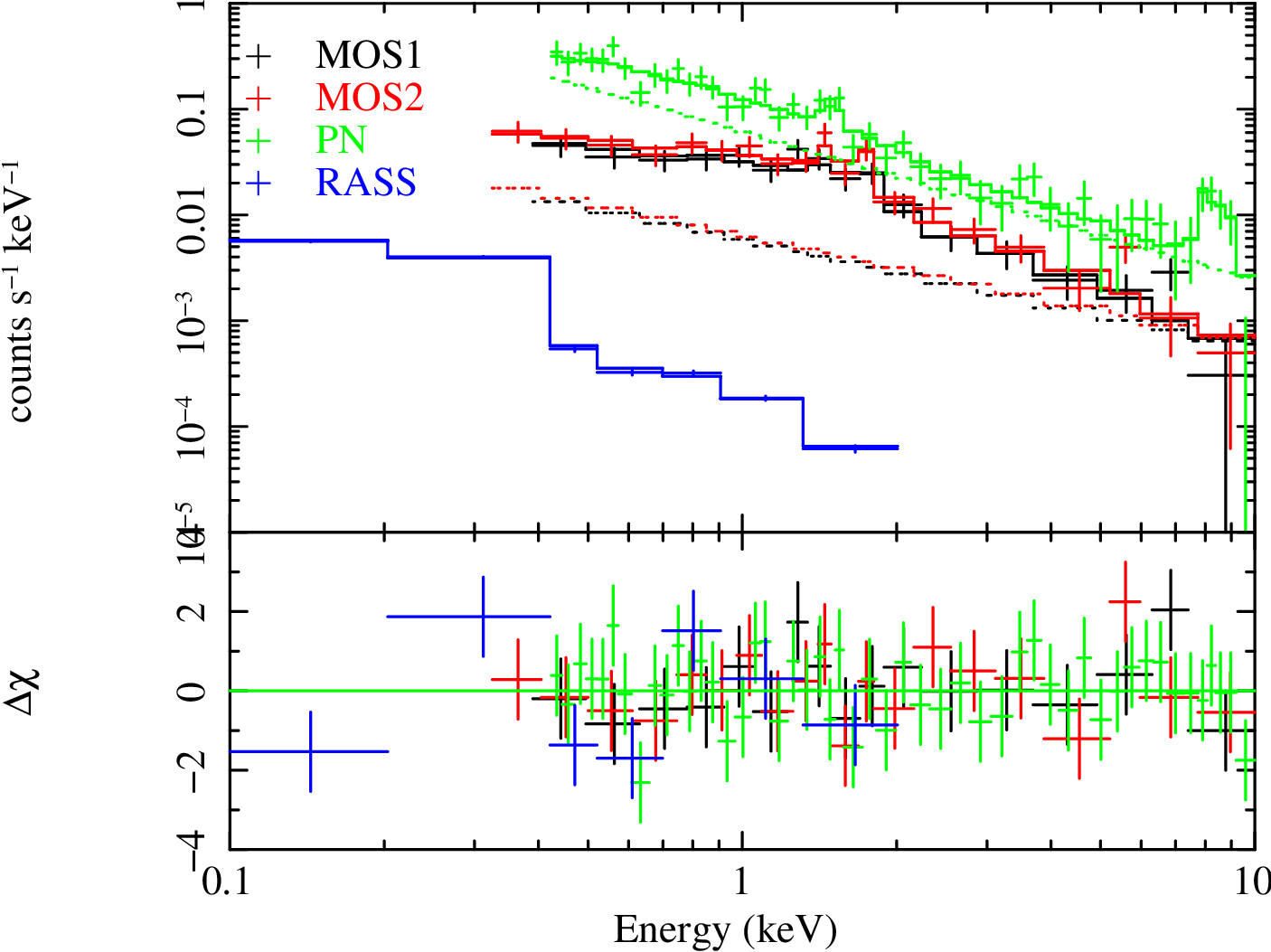}
 \end{center}
 \caption{Example of spectral fit. The upper panel shows the MOS1
   (black), MOS2 (red), and PN (green) spectra of HSC~J161136+541635
   at $z=0.332$ and the RASS background spectrum (blue).  The solid
   and dotted lines represent the total model consisting of cluster
   emission and backgrounds (see section~\ref{subsec:spec}) and the
   residual soft-proton background component, respectively. The
 lower panel shows the residual of the fit.}\label{fig:spec}
\end{figure}

The spectral model consists of (i) cluster thermal emission and (ii)
background components. For (i), we used the APEC thin-thermal plasma
model version 3.0.8 \citep{Smith01,Foster12} with the Galactic
photoelectric absorption model phabs \citep{Balucinska92}. The cluster
redshift and metal abundance were fixed at the optical value
[Table~\ref{tbl:sample}; \cite{Oguri18}] and at 0.3~solar,
respectively.  The Galactic hydrogen column density $N_{\rm H}$ was
fixed at a value taken from the Leiden/Argentine/Bonn survey
\citep{Kalberla05}. For (ii), the Galactic emission and the cosmic
X-ray background were evaluated by jointly fitting the RASS spectra
\citep{Snowden97} taken from the $0\degree.5-1\degree$ ring region
around the cluster. The other components due to possible solar wind
charge exchange, soft proton events, and instrumental fluorescent
lines were determined by adding a power-law model and narrow Gaussian
lines to the model.  An example of the spectral fitting is shown in
Figure~\ref{fig:spec}. The resultant APEC model parameters are
summarized in Table~\ref{tbl:spec}. The bolometric luminosity was
estimated from the best-fit model flux in the source-frame energy
range of 0.01 -- 30~keV. The missing flux due to the point-source
removal was corrected by interpolating the ICM emission assuming that
the observed cluster brightness profile is approximated by the
$\beta$-model \citep{Cavaliere97}.

The XMM + RASS joint fitting gives a reasonable result for most of
clusters; however, the background subtraction is not perfect at high
energies, particularly for the three clusters, HSC~J021115-034319,
HSC~J021427-062720, and HSC~J161039+540554. This is likely to be due
to the residual soft proton flares, as indicated by the count-rate
ratio between in-FOV and out-FOV \citep{DeLuca04}.  The EPIC-MOS1
(MOS2) count-rate ratio (in-FOV)/(out-FOV) is 2.7 (2.2), 1.1 (1.2),
1.4 (1.4) for HSC~J021115-034319, HSC~J021427-062720, and
HSC~J161039+540554, respectively. Thus, to check the background
uncertainty, we subtract the local background extracted from an
$r=(2-3)R_{500}$ annulus centered on the X-ray centroid and fit the
APEC model to the observed spectra. Since the resultant parameters are
consistent with those obtained from the XMM + RASS joint analysis
within that statistics for 14 clusters, we quote the values obtained
from the analysis by using the local background for the three clusters
mentioned above (see Table~\ref{tbl:spec}).

For 20 clusters with low counts, we convert the observed cluster
counts to bolometric luminosity assuming the APEC model with
temperature inferred by the $N-T$ relation \citep{Oguri18}. The
background was estimated from an $r=(2-3)R_{500}$ annulus centered on
the X-ray centroid. We checked the robustness of this method by
applying the same procedure to 17 clusters with higher counts. In most
of them, the luminosity agrees with the result of spectral analysis
within the $1\sigma$ statistical errors, while several clusters have a
larger uncertainty up to a factor of $\sim3$. Therefore, for 20
clusters, we take into account the upper limit on the luminosity when
we fit the $L-T$ relation (section~\ref{subsec:lt}).

\begin{table*}[hbt]
  \tbl{Results of spectral analysis under APEC thermal plasma model.}{%
  \begin{tabular}{lllll}\hline\hline
 	Cluster & $N_{\rm H}$ & $kT^{\mathrm{a}}$ & $L_X^{\mathrm{b}}$ & $\chi^2$/d.o.f. \\ 
	& ($10^{20}~{\rm cm^{-2}}$) & (keV) & ($10^{44}{\rm erg\,s^{-1}}$) & \\ \hline	
	 HSCJ142624-012657	&	3.21 	&	$	5.03 	_{	-1.10 	}^{+	1.18 	}	$	&	$	4.22 	_{	-0.53 	}^{+	0.41 	}	$	&	71.3 	/	63 	\\
 HSCJ021115-034319	&	1.95 	&	$	6.90 	_{	-2.43 	}^{+	6.16 	}	$	&	$	6.01 	_{	-1.47 	}^{+	2.00 	}	$	&	95.3 	/	81 	\\
 HSCJ095939+023044	&	1.71 	&	$	4.08 	_{	-0.69 	}^{+	0.76 	}	$	&	$	2.60 	_{	-0.46 	}^{+	0.43 	}	$	&	105.5 	/	108 	\\
 HSCJ161136+541635	&	0.95 	&	$	3.25 	_{	-0.75 	}^{+	1.27 	}	$	&	$	1.89 	_{	-0.34 	}^{+	0.55 	}	$	&	73.8 	/	69 	\\
 HSCJ090914-001220	&	2.73 	&	$	4.07 	_{	-0.98 	}^{+	1.26 	}	$	&	$	1.78 	_{	-0.40 	}^{+	0.30 	}	$	&	48.4 	/	33 	\\
 HSCJ141508-002936	&	3.27 	&	$	2.14 	_{	-0.18 	}^{+	0.17 	}	$	&	$	0.58 	_{	-0.06 	}^{+	0.05 	}	$	&	471.1 	/	404 	\\
 HSCJ140309-001833	&	3.66 	&	$	2.96 	_{	-0.62 	}^{+	0.70 	}	$	&	$	1.93 	_{	-0.42 	}^{+	0.25 	}	$	&	114.4 	/	122 	\\
 HSCJ095737+023426	&	1.85 	&	$	3.30 	_{	-0.35 	}^{+	0.47 	}	$	&	$	2.58 	_{	-0.27 	}^{+	0.27 	}	$	&	153.6 	/	154 	\\
 HSCJ022135-062618	&	2.73 	&	$	5.66 	_{	-3.66 	}^{+	6.82 	}	$	&	$	0.60 	_{	-0.50 	}^{+	1.15 	}	$	&	23.9 	/	13 	\\
 HSCJ232924-004855	&	4.33 	&	$	4.80 	_{	-1.44 	}^{+	5.02 	}	$	&	$	0.91 	_{	-0.14 	}^{+	0.43 	}	$	&	60.7 	/	38 	\\
 HSCJ022512-062259	&	2.95 	&	$	2.09 	_{	-0.40 	}^{+	0.61 	}	$	&	$	0.76 	_{	-0.18 	}^{+	0.24 	}	$	&	78.8 	/	71 	\\
 HSCJ021427-062720	&	2.13 	&	$	4.76 	_{	-0.82 	}^{+	1.18 	}	$	&	$	1.39 	_{	-0.18 	}^{+	0.19 	}	$	&	62.7 	/	62 	\\
 HSCJ161039+540554	&	0.94 	&	$	1.23 	_{	-0.29 	}^{+	0.86 	}	$	&	$	0.52 	_{	-0.12 	}^{+	0.10 	}	$	&	45.6 	/	43 	\\
 HSCJ095903+025545	&	1.79 	&	$	1.64 	_{	-0.30 	}^{+	0.36 	}	$	&	$	0.48 	_{	-0.21 	}^{+	0.35 	}	$	&	126.7 	/	160 	\\
 HSCJ100049+013820	&	1.80 	&	$	3.28 	_{	-1.44 	}^{+	2.47 	}	$	&	$	0.78 	_{	-0.36 	}^{+	0.56 	}	$	&	114.9 	/	119 	\\
 HSCJ090743+013330	&	3.20 	&	$	1.08 	_{	-0.12 	}^{+	0.92 	}	$	&	$	0.21 	_{	-0.05 	}^{+	0.28 	}	$	&	31.0 	/	22 	\\
 HSCJ095824+024916	&	1.84 	&	$	2.03 	_{	-0.35 	}^{+	0.60 	}	$	&	$	0.17 	_{	-0.04 	}^{+	0.05 	}	$	&	314.9 	/	280 	\\\hline
HSCJ090754+005732	&	3.04	&		3.6							&	$	3.30 	\pm	0.69 				$	&				\\
HSCJ090541+013226	&	3.51	&		3.4							&	$				<	0.63 		$	&				\\
HSCJ232924-004855	&	4.33	&		3.3							&	$				<	0.29 		$	&				\\
HSCJ222210-004421	&	4.72	&		3.3							&	$	0.23 	\pm	0.49 				$	&				\\
HSCJ100221+032807	&	1.66	&		3.1							&	$				<	1.24 		$	&				\\
HSCJ221211-000821	&	4.40	&		3.1							&	$	0.92 	\pm	0.34 				$	&				\\
HSCJ160424+430438	&	1.09	&		3.0							&	$				<	0.19 		$	&				\\
HSCJ100300+013152	&	2.07	&		3.0							&	$	0.66 	\pm	0.23 				$	&				\\
HSCJ142203-000402	&	2.86	&		3.0							&	$				<	0.09 		$	&				\\
HSCJ090419+020641	&	3.66	&		3.0							&	$				<	0.33 		$	&				\\
HSCJ220625+013905	&	4.17	&		3.0							&	$	1.04 	\pm	0.63 				$	&				\\
HSCJ221422+004706	&	3.50	&		2.9							&	$				<	0.21 		$	&				\\
HSCJ090806+011956	&	3.12	&		2.8							&	$	1.18 	\pm	0.12 				$	&				\\
HSCJ090509+012428	&	3.59	&		2.8							&	$				<	0.51 		$	&				\\
HSCJ022246-061703	&	2.78	&		2.8							&	$				<	0.35 		$	&				\\
HSCJ222121-004630	&	4.73	&		2.7							&	$	2.18 	\pm	0.76 				$	&				\\
HSCJ221726-001020	&	4.78	&		2.7							&	$				<	0.17 		$	&				\\
HSCJ090602+011443	&	3.44	&		2.6							&	$				<	0.15 		$	&				\\
HSCJ221538+004227	&	3.71	&		2.5							&	$				<	0.47 		$	&				\\
HSCJ141648+521039	&	1.07	&		2.5							&	$	0.06 	\pm	0.13 				$	&				\\
 \hline
  \end{tabular}}\label{tbl:spec}
  \begin{tabnote}
    $^{\mathrm{a}}$ { The gas temperature in keV. For 17 high-count
      clusters, $kT$ was derived from spectral fitting. For 20
      low-counts cluters, $kT$ was estimated from richness and the
      $N-T$ relation.}  $^{\mathrm{b}}$ The bolometric luminosity
    within the scale radius $R_{500}$
  \end{tabnote}
\end{table*}

\section{Results}\label{sec:results}

\subsection{Centroid offset and peak offset}\label{subsec:centroidoffset}
\begin{figure*}[htb]
 \begin{center}
  \includegraphics[scale=0.5,angle=90]{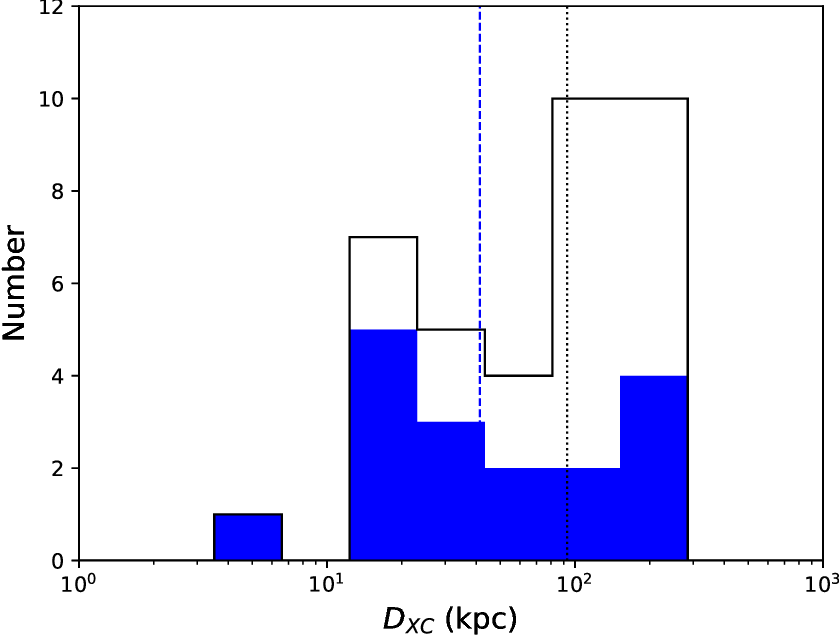}
  \includegraphics[scale=0.5,angle=90]{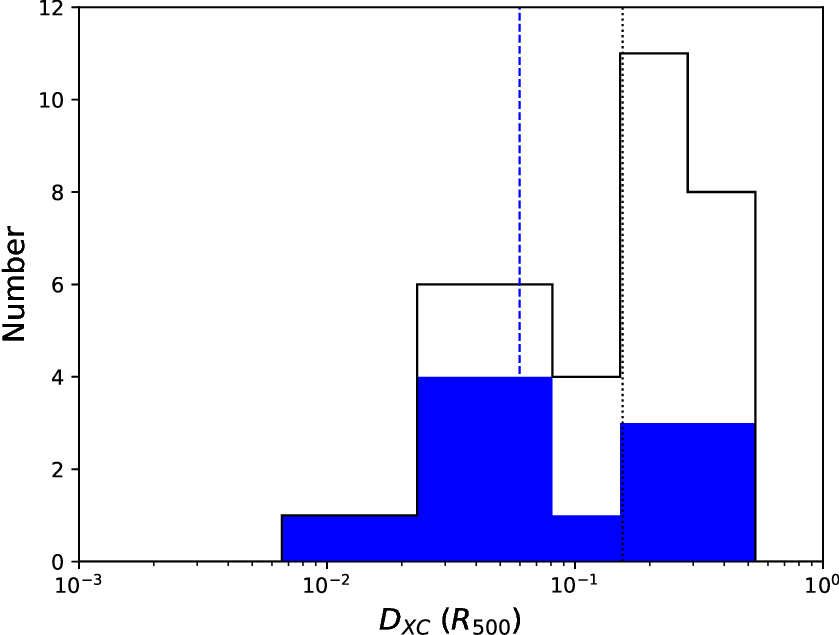}

 \includegraphics[scale=0.5,angle=90]{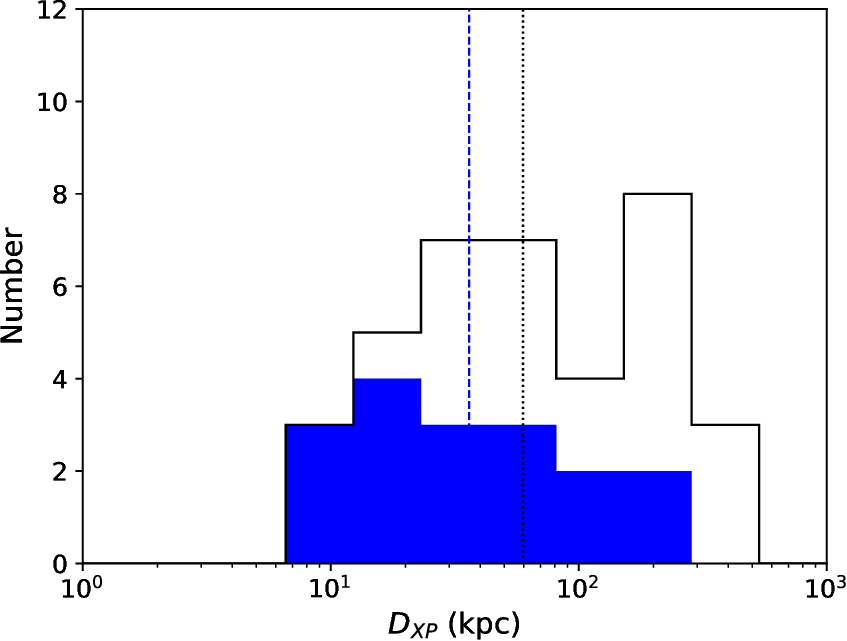}
 \includegraphics[scale=0.5,angle=90]{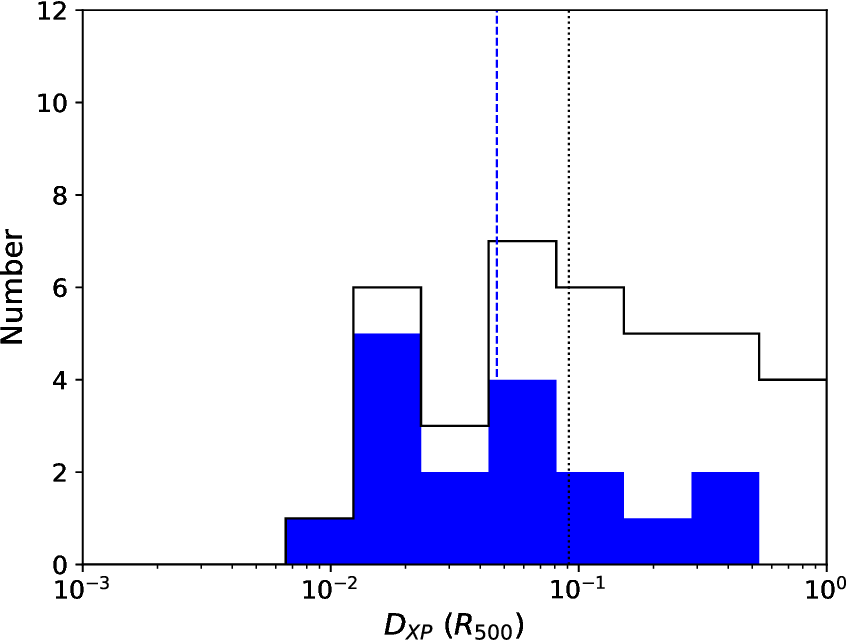}
 \end{center}
 \caption{Histograms of centroid offset (upper panels) and peak offset
   (lower panels) in units of kpc (left panels) and $R_{500}$ (right
   panels). The filled and open histograms show the distributions of
   17 high-count clusters and all clusters. In each panel, the
   vertical dashed line indicates the distribution
   median. }\label{fig:centroid}
\end{figure*}

We define the centroid offset $D_{\rm XC}$ as a projected distance
between the BCG coordinates and the X-ray centroid measured within
$R_{500}$. The measured centroid offset is given in
Table~\ref{tbl:sample}. The histograms of the centroid offset in kpc
and fractions of $R_{500}$ are shown in the upper panels of
Figure~\ref{fig:centroid}. The median values of centroid offset are
given in Table~\ref{tbl:centroidoffset}, which is $\tilde{D}_{\rm
  XC}=41$~kpc or $0.06R_{500}$ for 17 clusters and $\tilde{D}_{\rm
  XC}=92$~kpc or $0.16R_{500}$ for the entire sample.

\begin{table}[hbt]
  \tbl{Median values of centroid offset and peak offset.}{%
  \begin{tabular}{llllll}\hline\hline
              & $D_{\rm XC}$ & $D_{\rm XC}$ & $D_{\rm XP}$ & $D_{\rm XP}$ & Fraction$^{\mathrm{a}}$ \\ 
              & (kpc) & ($R_{500}$) & (kpc) & ($R_{500}$) &   \\ \hline
  17 clusters & 41  & 0.06 & 36  & 0.05 & $29\pm 11 (\pm 13)$ \% \\ 
  20 clusters & 112 & 0.18 & 130 & 0.22 & $(< 5)$ \%  \\
  All         & 92  & 0.16 & 56  & 0.29 & $14\pm 6 (\pm 6)$ \% \\
 \hline
  \end{tabular}}\label{tbl:centroidoffset}
 \begin{tabnote}
 $^{\mathrm{a}}$ Fraction of relaxed clusters and the statistical (systematic) errors (see text).
 \end{tabnote}
\end{table}

Next, we measured the X-ray peak position within $R_{500}$ by using
the XMM composite image smoothed with a $\sigma=3$~(pixels) Gaussian
function. We define the peak offset $D_{\rm XP}$ as a projected
distance relative to the BCG coordinates. The resultant peak offset is
shown in Table~\ref{tbl:sample} and the median values are given in
Table~\ref{tbl:centroidoffset}.  The lower panels of
Figure~\ref{fig:centroid} show the histograms of the measured peak
offset in units of kpc and $R_{500}$. For 17 clusters, the median is
$\tilde{D}_{\rm XP}=36$~kpc or $0.05R_{500}$. For the entire sample,
$\tilde{D}_{\rm XP}=56$~kpc or $0.29R_{500}$.

The twenty clusters with low statistics tend to show larger centroid
and peak offsets. As discussed in \cite{Mann12}, the accuracy of the
X-ray peak position depends on the statistical quality of the X-ray
observations as well as the surface brightness distribution, which
varies significantly between clusters.  We assessed the standard error
of the peak offset $\delta D_{\rm XP}$ by comparing X-ray images of
each cluster with different smoothing scale
($\sigma=2,3,4$~pixels). For 17 clusters, $\delta D_{\rm XP}$ ranges
from 3\% to 68\% (the mean is 24\%). On the other hand, the entire
sample contains objects with larger uncertainties, resulting
$2\%<\delta D_{\rm XP}< 113\%$ (the mean is 22\%).

We divide the sample into two classes, ``relaxed'' clusters with a
small peak offset ($D_{\rm XP}<0.02R_{500}$) and ``disturbed''
clusters with a large offset ($D_{\rm XP}>0.02R_{500}$) following the
criteria used in \cite{Sanders09}. As a result, there are only 5
relaxed clusters and the fraction of relaxed objects is $14\pm 6 (\pm
6)\%$ for the entire sample, and $29\pm 11 (\pm 13)$\% if 20 clusters
with low photon statistics are excluded.  Here the first error is the
statistical uncertainty estimated by the bootstrap resampling method
\citep{Efron82} and second error in the parenthesis indicates the
systematic uncertainty in the measurement and was estimated by
referring to the peak-to-peak amplitude of the relaxed fraction in the
case that we varied the smoothing scale of the X-ray images between 2
and 4~pixels.  Section~\ref{subsec:discussion_centroid} compares the
fraction of relaxed clusters in the optical clusters with nearby X-ray
and SZE cluster samples.

\subsection{Luminosity-temperature relation}\label{subsec:lt}
\begin{figure*}[htb]
 \begin{center}
  \includegraphics[scale=0.5, angle=90]{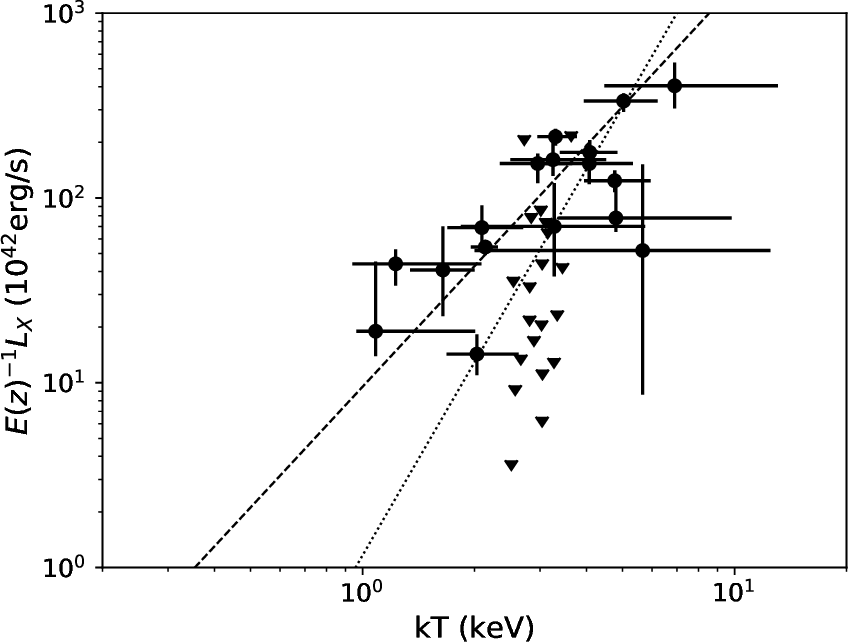}
  \includegraphics[scale=0.5, angle=90]{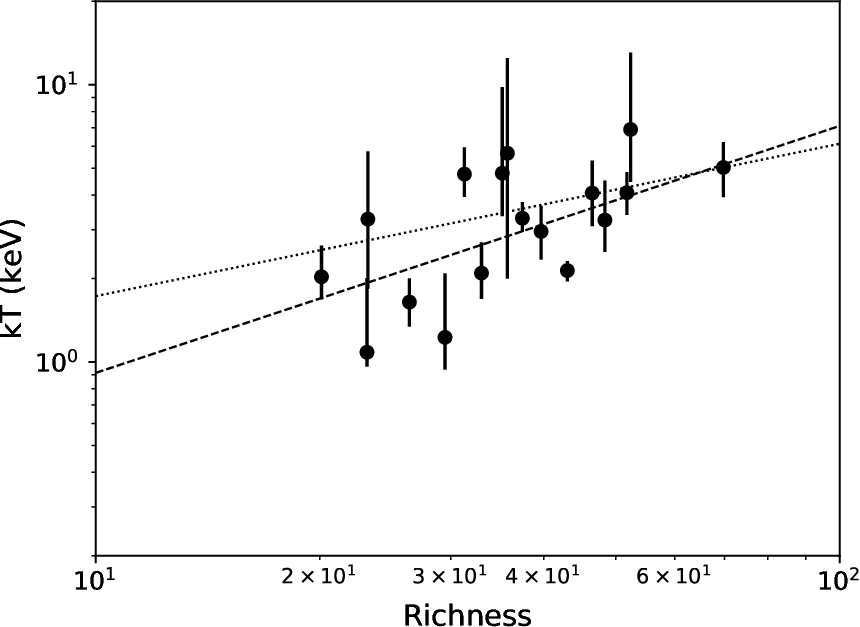}
 \end{center}
 \caption{(left) Luminosity-temperature relation of the high-richness
   clusters.  (right) Gas temperature - richness relation. In each
   panel, the circles are 17 clusters with high photon statistics and
   the dashed line shows the best-fit power-law model. In the left
   panel, the triangles are 20 clusters with low statistics and the
   dotted line indicates the best-fit model derived for the entire
   sample.  In the right panel, the dotted line shows the best-fit
   $T-\hat{N}_{\rm mem}$ relation derived for the XXL and XXL-LSS
   sample \citep{Oguri18}.  }\label{fig:lt}
\end{figure*}

In the self-similar model, the redshift evolution of the cluster
scaling relations is described by the factor $E(z) = (\Omega_M(1+z)^3
+ \Omega_{\Lambda})^{1/2}$ and the luminosity of the cluster gas in
the hydrostatic state follows $E(z)^{-1}L\propto T^2$. Within this
framework, the normalization of the luminosity-temperature relation
evolves as $E(z)^{\gamma}$ \citep{Giles16}. Despite a number of
observational studies, however, no clear consensus has been reached on
the evolution of the scaling relations \citep[for a review,
see][]{Giodini13}. In the present paper, we correct the redshift
evolution by applying the self-similar model and plot $E(z)^{-1}L$
against gas temperature in the left panel of Figure~\ref{fig:lt}.

We fit the observed $L_X-T$ relation to the power-law model
(equation~\ref{eq:model_lt}). To account for measurement errors in
both variables, we use the Bayesian regression method \citep{Kelly07}
because it has been demonstrated that it outperforms other common
estimators that can constrain the parameters even when data have large
measurement errors or only upper limits.  The quantities $a$, $b$, and
the intrinsic scatter are treated as free parameters.
\begin{equation}
\log{\left( \frac{E(z)^{-1}L_X}{10^{42}{\rm erg\,s^{-1}}}\right ) } = a + b \log{\left(\frac{T}{\rm keV}\right)} \label{eq:model_lt}
\end{equation}
For 17 clusters, the best-fit parameters are $a=0.98\pm 0.29$,
$b=2.17\pm 0.61$, and $\sigma_{L|T}=0.20\pm0.09$. For the entire
sample, the uncertainties become larger; $a=0.07\pm 0.70$, $b=3.47\pm
1.44$, and $\sigma_{L|T}=0.43\pm0.06$, however, they are consistent
within the statistical errors.

For comparison, if we apply the BCES code \citep{Akritas96} to the
present optical sample, the fitting yields the best-fit $L_X-T$ slope
steeper than 2.0 but with a fairly large uncertainty; namely,
$b=2.59\pm3.31$. \cite{Kelly07} noted that the BCES estimate of the
slope tends to suffer some bias and becomes considerably unstable when
the measurement errors are large and/or the sample size is
small. Therefore, in section~\ref{subsec:discussion_scaling} we quote
the above results based on the Bayesian regression method.

Since the present sample was selected by cross-matching the optical
clusters with the X-ray catalog and their exposures available in the
XMM-Newton archive is not homogeneous, we study the impact of
selection effect on the scaling relation as follows: i) 37 clusters
are randomly chosen out of the CAMIRA catalog, ii) the X-ray
temperature and luminosity are estimated from the richness assuming
the best-fit $N-T$ and $L-T$ relations and the intrinsic scatters, and
the $1\sigma$ errors are assigned. Here only upper limit on luminosity
is given to 12 clusters to mimic the actual observations
(Table~\ref{tbl:spec}). iii) the $L-T$ relation was fit to determine
the coefficients in equation~\ref{eq:model_lt}.  iv) steps i)--iii)
are repeated $10^3$ times. From the resultant parameter distributions,
we find that $b$ ($a$) tends to be underestimated (overestimated) by
$\Delta b=0.23$ ($\Delta a=0.15$).  A similar trend was seen if we
limit the sample to 17 clusters with better statistics. We thus regard
the above quantities as systematic errors of the $L_X-T$ relation
caused by the selection effect and compare our results with previous
studies in section~\ref{subsec:discussion_scaling}.

\section{Discussion}\label{sec:discussion}
Our analyses of two sub-samples yielded consistent results within
measurement uncertainties. Due to the shallow exposures of 20
clusters, however, we will discuss interpretations of the present
results primarily based on the observations of 17 clusters with higher
counts.

\subsection{Centroid offset and the cluster dynamical state}\label{subsec:discussion_centroid}
In section~\ref{subsec:centroidoffset} we quantified the centroid
offset and peak offset from the XMM image analysis to find that half
of the sample has the centroid (peak) offset larger than $0.06R_{500}$
($0.05R_{500}$) or 41~kpc (36~kpc). Following the criteria used in
\cite{Sanders09}, \cite{Rossetti16} estimated the fraction of relaxed
clusters in the Planck SZE sample to be $52\pm4$\%. They also
calculated the fraction to be $\sim 74$\% in X-ray selected cluster
samples constructed from the HIFLUGCS, MACS, and REXCESS surveys,
whereas we obtain only $29\pm11 (\pm 13)$\% from our optical
sample. This suggests that the optical cluster sample contains a
larger fraction of disturbed clusters particularly in comparison with
the X-ray selected cluster samples.

X-ray observations preferentially detect relaxed clusters having cool
cores at the center as opposed to more disturbed, non-cool-core
clusters found in SZE surveys \citep{Eckert11,
  Rossetti17,Andrade-Santos17, Lovisari17}.  \cite{Andreon16} found a
wider population in the sample selected independently of X-ray
properties than seen in the SZE surveys.  Furthermore, \cite{Chon17}
claim that the cool-core bias in previous X-ray surveys is due to the
survey-selection method such as for a flux-limited survey, and is not
due to the inherent nature of X-ray selection. Therefore, considering
the nature of the HSC cluster survey, we suggest that the observed
small fraction of relaxed clusters in the present optical sample is
due to the fact that the CAMIRA algorithm is immune to the dynamical
state of X-ray-emitting gas and is likely to detect clusters with a
wider range of cluster morphology. This needs to be confirmed with the
large uniform sample to be constructed by the eROSITA survey.

Given a higher merger rate in the distant universe, the redshift
evolution of X-ray morphology is likely to affect the measurement of
the fraction of relaxed clusters with respect to disturbed clusters.
\cite{Mann12} reported based on the Chandra observations that the
fraction of morphologically-disturbed clusters increases at $z>0.4$
for the X-ray luminous clusters. On the other hand, \cite{McDonald17}
found that there is no measurable redshift evolution in the X-ray
morphology of massive clusters.　In our optical cluster sample, the
redshift evolution is not significantly seen; the fractions of relaxed
clusters estimated from the X-ray peak offsets are $38\pm13$\% at
$z<0.4$ and $<25$\% at $z>0.4$.

\subsection{Scaling relations}\label{subsec:discussion_scaling}
We obtained the slope of $2.2 \pm0.6 (\pm 0.2)$ of the $L_X-T$
relation for the present optical clusters.
% (section~\ref{subsec:lt}).
Note that the second error in the parenthesis indicates the systematic
uncertainty due to the sample selection
(section~\ref{subsec:lt}). This is consistent with the slope of 2.0
predicted from the self-similar model, whereas a steeper slope of
$\sim 3$ has been reported by many X-ray observations in the past
\citep[for review, see][]{Giodini13}. Even so, the data points lie
within the observed large scatter of X-ray clusters on the $L_X-T$
plane \citep{Takey11}.

The gas density in the core region is known to have a significant
scatter, and the self-similar relation is not satisfied particularly
in clusters with a compact, cooling core \citep[e.g.,][]{Ota06}. In
the present sample, the fraction of relaxed cluster is $\sim30$\%
(section~\ref{subsec:centroidoffset}) and our preliminary analysis of
the X-ray brightness profile shows that there are only a few objects
that have a very small core radius ($r_c<100$~kpc), suggesting that
the impact of cool core on the $ L_{X}-T$ relation is likely to be
small.

The fitted slope agrees with that of the Red-sequence Cluster Survey
at high redshifts [the slope parameter is $2.1\pm0.3$; \cite{Hicks08}]
and that of the total RCS sample; namely, 18 clusters at $0.16<z<1.0$
[$2.7\pm0.5$; \cite{Hicks13}] within the errors. In comparison with
X-ray selected samples that contain a large number of clusters
($>100$) at a wide redshift range [$2.53\pm0.15$; \cite{Reichert11},
$2.80\pm0.12$; \cite{Takey11}, $2.72\pm0.18$; \cite{Maughan12}], the
present sample shows a marginally shallower slope. To further confirm
the result, however, we need to increase the number of clusters and
improve the accuracy with which the $L_X-T$ relation is measured.

The right panel of Figure~\ref{fig:lt} shows the relationship between
gas temperature and optical richness. Although the scatter is large,
the positive correlation is seen and the correlation coefficient is
calculated to be 0.58. Assuming the power-law model,
\begin{equation}
\log{\left(\frac{T}{\rm keV}\right)}  = a_T \log{\left(\frac{\hat{N}_{\rm mem}}{30}\right)} + b_T, \label{eq:model_tn}
\end{equation}
the fit to the data yields $a_T = 0.89\pm0.33$ and $b_T=0.39\pm0.06$.
This is marginally steeper than the best-fit power-law relation
derived for 50 bright X-ray clusters in the XXL and XXL-LSS fields
[$a_T=0.50\pm0.12$, $b_T=0.48\pm0.02$; \cite{Oguri18}].  Because the
gas temperature of XXL and XXL-LSS clusters was measured in the
central $r<300$~kpc region \citep{Pierre04,Pierre16}, direct
comparison is not easy.  Conversely, the self-similar model predicts
$T\propto \hat{N}_{\rm mem}^{2/3}$ given that the cluster mass is
related to richness and temperature through $M\propto \hat{N}_{\rm
  mem}$ and $M\propto T^{3/2}$, respectively. Thus our fitting result
is consistent with the self-similar model, although the statistical
uncertainty is large.  Recently, \cite{Okabe18} reported based on the
weak-lensing analysis of 1750 clusters in the CAMIRA catalog
\citep{Oguri18} that the $M-N$ relation has a stepper slope of $M
\propto \hat{N}_{\rm mem}^{1.41\pm0.24}$. This modifies the above
expectation from $T \propto \hat{N}_{\rm mem}^{2/3}$ to $T\propto
\hat{N}_{\rm mem}N^{0.94\pm0.16}$, which is, however, consistent with
our result within the $1\sigma$ error.

\section{Summary and future prospects}\label{sec:summary}
Using the XMM-Newton archive data, we apply an X-ray analysis to 37
rich, optical clusters of galaxies at $0.1<z<1.1$ in the HSC-SSP
field. Most of the clusters were serendipitously detected in the
XMM-Newton fields of view. We subdivided the sample to two, i.e., 17
(20) clusters with high (low) photon counts, to find the results of
two subsamples agree with each other within errors. Due to large
statistical uncertainties in the 20 clusters, however, we discussed
the implications mainly based on the results for 17 clusters with
sufficient statistics. The major findings are as follows:
\begin{enumerate}
\item We systematically analyzed the X-ray centroid or peak offset as
  compared with the BCG position. The fraction of relaxed clusters in
  the optical cluster sample, which is defined based on the offset
  between the BCG and X-ray peak, is $29\pm11(\pm 13)$\%. This is less
  than that of the X-ray samples. Because the optical sample is immune
  to the cool-core bias, it is likely to contain more disturbed
  clusters and thus cover a larger range of the cluster morphology.
\item The slope of the luminosity-temperature relation is marginally
  less than that of X-ray samples and is consistent with the
  self-similar model prediction of 2.0. The slope of the
  temperature-richness relation is also consistent with the prediction
  of the self-similar model although the former has a large
  statistical uncertainty.
\end{enumerate}

Our pilot study provides important information about the X-ray
properties of the optical clusters, which are marginally different
from those observed in the X-ray samples. To obtain more conclusive
results, we need to improve the measurement accuracy and the sample
uniformity. We thus plan to extend the analysis by (1) incorporating
fainter objects in the 3MM-DR7 catalog and (2) conducting X-ray
observations of the massive, high-redshift ($0.8<z<1.2$) clusters
newly discovered by the HSC-SSP survey. For the latter, the XMM-Newton
follow-up project is now ongoing and is to be the subject of an
upcoming presentation.  Furthermore, by the time of completion of the
HSC-SSP survey, the CAMIRA cluster catalog will be about 6-times
larger than that at present.  These works should allow us to derive
the mass-observable scaling by using a larger number of clusters and
study the redshift evolution of the X-ray properties of the optical
clusters. Detailed comparisons of optical, weak lensing, SZE, and
X-ray selected clusters will improve our knowledge of cluster-mass
calibration and cluster evolution.

\begin{ack}
  The Hyper Suprime-Cam (HSC) collaboration includes the astronomical
  communities of Japan and Taiwan, and Princeton University. The HSC
  instrumentation and software were developed by the National
  Astronomical Observatory of Japan (NAOJ), the Kavli Institute for
  the Physics and Mathematics of the Universe (Kavli IPMU), the
  University of Tokyo, the High Energy Accelerator Research
  Organization (KEK), the Academia Sinica Institute for Astronomy and
  Astrophysics in Taiwan (ASIAA), and Princeton University. Funding
  was contributed by the FIRST program from Japanese Cabinet Office,
  the Ministry of Education, Culture, Sports, Science and Technology
  (MEXT), the Japan Society for the Promotion of Science (JSPS), Japan
  Science and Technology Agency (JST), the Toray Science Foundation,
  NAOJ, Kavli IPMU, KEK, ASIAA, and Princeton University.

  This paper makes use of software developed for the Large Synoptic
  Survey Telescope. We thank the LSST Project for making their code
  available as free software at http://dm.lsst.org

  The Pan-STARRS1 Surveys (PS1) have been made possible through
  contributions of the Institute for Astronomy, the University of
  Hawaii, the Pan-STARRS Project Office, the Max-Planck Society and
  its participating institutes, the Max Planck Institute for
  Astronomy, Heidelberg and the Max Planck Institute for
  Extraterrestrial Physics, Garching, The Johns Hopkins University,
  Durham University, the University of Edinburgh, Queen's University
  Belfast, the Harvard-Smithsonian Center for Astrophysics, the Las
  Cumbres Observatory Global Telescope Network Incorporated, the
  National Central University of Taiwan, the Space Telescope Science
  Institute, the National Aeronautics and Space Administration under
  Grant No. NNX08AR22G issued through the Planetary Science Division
  of the NASA Science Mission Directorate, the National Science
  Foundation under Grant No. AST-1238877, the University of Maryland,
  and Eotvos Lorand University (ELTE) and the Los Alamos National
  Laboratory.

  We are grateful to Chien-Hsiu Lee for useful comments.  This work
  was supported in part by JSPS KAKENHI grants 16K05295 (NO) and
  JP15K17610 (SU).  YI is supported by Rikkyo University Special Fund
  for Research (SFR).  This work is supported in part by the Ministry
  of Science and Technology of Taiwan (grant MOST
  106-2628-M-001-003-MY3) and by Academia Sinica (grant
  AS-IA-107-M01). We thank the anonymous referee for useful
  suggestions.
\end{ack}

\bibliographystyle{apj}
\bibliography{ref}

\begin{thebibliography}{75}
\expandafter\ifx\csname natexlab\endcsname\relax\def\natexlab#1{#1}\fi

\bibitem[{{Aihara} {et~al.}(2018{\natexlab{a}}){Aihara}, {Armstrong},
  {Bickerton}, {Bosch}, {Coupon}, {Furusawa}, {Hayashi}, {Ikeda}, {Kamata},
  {Karoji}, {Kawanomoto}, {Koike}, {Komiyama}, {Lang}, {Lupton}, {Mineo},
  {Miyatake}, {Miyazaki}, {Morokuma}, {Obuchi}, {Oishi}, {Okura}, {Price},
  {Takata}, {Tanaka}, {Tanaka}, {Tanaka}, {Uchida}, {Uraguchi}, {Utsumi},
  {Wang}, {Yamada}, {Yamanoi}, {Yasuda}, {Arimoto}, {Chiba}, {Finet},
  {Fujimori}, {Fujimoto}, {Furusawa}, {Goto}, {Goulding}, {Gunn}, {Harikane},
  {Hattori}, {Hayashi}, {He{\l}miniak}, {Higuchi}, {Hikage}, {Ho}, {Hsieh},
  {Huang}, {Huang}, {Imanishi}, {Iwata}, {Jaelani}, {Jian}, {Kashikawa},
  {Katayama}, {Kojima}, {Konno}, {Koshida}, {Kusakabe}, {Leauthaud}, {Lee},
  {Lin}, {Lin}, {Mandelbaum}, {Matsuoka}, {Medezinski}, {Miyama}, {Momose},
  {More}, {More}, {Mukae}, {Murata}, {Murayama}, {Nagao}, {Nakata}, {Niida},
  {Niikura}, {Nishizawa}, {Oguri}, {Okabe}, {Ono}, {Onodera}, {Onoue}, {Ouchi},
  {Pyo}, {Shibuya}, {Shimasaku}, {Simet}, {Speagle}, {Spergel}, {Strauss},
  {Sugahara}, {Sugiyama}, {Suto}, {Suzuki}, {Tait}, {Takada}, {Terai}, {Toba},
  {Turner}, {Uchiyama}, {Umetsu}, {Urata}, {Usuda}, {Yeh}, \&
  {Yuma}}]{HSC1styr}
{Aihara}, H., {Armstrong}, R., {Bickerton}, S., {et~al.} 2018{\natexlab{a}},
  \pasj, 70, S8

\bibitem[{{Aihara} {et~al.}(2018{\natexlab{b}}){Aihara}, {Arimoto},
  {Armstrong}, {Arnouts}, {Bahcall}, {Bickerton}, {Bosch}, {Bundy}, {Capak},
  {Chan}, {Chiba}, {Coupon}, {Egami}, {Enoki}, {Finet}, {Fujimori}, {Fujimoto},
  {Furusawa}, {Furusawa}, {Goto}, {Goulding}, {Greco}, {Greene}, {Gunn},
  {Hamana}, {Harikane}, {Hashimoto}, {Hattori}, {Hayashi}, {Hayashi},
  {He{\l}miniak}, {Higuchi}, {Hikage}, {Ho}, {Hsieh}, {Huang}, {Huang},
  {Ikeda}, {Imanishi}, {Inoue}, {Iwasawa}, {Iwata}, {Jaelani}, {Jian},
  {Kamata}, {Karoji}, {Kashikawa}, {Katayama}, {Kawanomoto}, {Kayo}, {Koda},
  {Koike}, {Kojima}, {Komiyama}, {Konno}, {Koshida}, {Koyama}, {Kusakabe},
  {Leauthaud}, {Lee}, {Lin}, {Lin}, {Lupton}, {Mandelbaum}, {Matsuoka},
  {Medezinski}, {Mineo}, {Miyama}, {Miyatake}, {Miyazaki}, {Momose}, {More},
  {More}, {Moritani}, {Moriya}, {Morokuma}, {Mukae}, {Murata}, {Murayama},
  {Nagao}, {Nakata}, {Niida}, {Niikura}, {Nishizawa}, {Obuchi}, {Oguri},
  {Oishi}, {Okabe}, {Okamoto}, {Okura}, {Ono}, {Onodera}, {Onoue}, {Osato},
  {Ouchi}, {Price}, {Pyo}, {Sako}, {Sawicki}, {Shibuya}, {Shimasaku},
  {Shimono}, {Shirasaki}, {Silverman}, {Simet}, {Speagle}, {Spergel},
  {Strauss}, {Sugahara}, {Sugiyama}, {Suto}, {Suyu}, {Suzuki}, {Tait},
  {Takada}, {Takata}, {Tamura}, {Tanaka}, {Tanaka}, {Tanaka}, {Tanaka},
  {Terai}, {Terashima}, {Toba}, {Tominaga}, {Toshikawa}, {Turner}, {Uchida},
  {Uchiyama}, {Umetsu}, {Uraguchi}, {Urata}, {Usuda}, {Utsumi}, {Wang}, {Wang},
  {Wong}, {Yabe}, {Yamada}, {Yamanoi}, {Yasuda}, {Yeh}, {Yonehara}, \&
  {Yuma}}]{HSC1styrOverview}
{Aihara}, H., {Arimoto}, N., {Armstrong}, R., {et~al.} 2018{\natexlab{b}},
  \pasj, 70, S4

\bibitem[{{Akritas} \& {Bershady}(1996)}]{Akritas96}
{Akritas}, M.~G., \& {Bershady}, M.~A. 1996, \apj, 470, 706

\bibitem[{{Allen} {et~al.}(2011){Allen}, {Evrard}, \& {Mantz}}]{Allen11}
{Allen}, S.~W., {Evrard}, A.~E., \& {Mantz}, A.~B. 2011, \araa, 49, 409

\bibitem[{{Andrade-Santos} {et~al.}(2017){Andrade-Santos}, {Jones}, {Forman},
  {Lovisari}, {Vikhlinin}, {van Weeren}, {Murray}, {Arnaud}, {Pratt},
  {D{\'e}mocl{\`e}s}, {Kraft}, {Mazzotta}, {B{\"o}hringer}, {Chon},
  {Giacintucci}, {Clarke}, {Borgani}, {David}, {Douspis}, {Pointecouteau},
  {Dahle}, {Brown}, {Aghanim}, \& {Rasia}}]{Andrade-Santos17}
{Andrade-Santos}, F., {Jones}, C., {Forman}, W.~R., {et~al.} 2017, \apj, 843,
  76

\bibitem[{{Andreon} {et~al.}(2016){Andreon}, {Serra}, {Moretti}, \&
  {Trinchieri}}]{Andreon16}
{Andreon}, S., {Serra}, A.~L., {Moretti}, A., \& {Trinchieri}, G. 2016, \aap,
  585, A147

\bibitem[{{Arnaud}(1996)}]{Arnaud96}
{Arnaud}, K.~A. 1996, in Astronomical Society of the Pacific Conference Series,
  Vol. 101, Astronomical Data Analysis Software and Systems V, ed. G.~H.
  {Jacoby} \& J.~{Barnes}, 17

\bibitem[{{Arnaud} {et~al.}(2005){Arnaud}, {Pointecouteau}, \&
  {Pratt}}]{Arnaud05}
{Arnaud}, M., {Pointecouteau}, E., \& {Pratt}, G.~W. 2005, \aap, 441, 893

\bibitem[{{Balucinska-Church} \& {McCammon}(1992)}]{Balucinska92}
{Balucinska-Church}, M., \& {McCammon}, D. 1992, \apj, 400, 699

\bibitem[{{Bleem} {et~al.}(2015){Bleem}, {Stalder}, {de Haan}, {Aird}, {Allen},
  {Applegate}, {Ashby}, {Bautz}, {Bayliss}, {Benson}, {Bocquet}, {Brodwin},
  {Carlstrom}, {Chang}, {Chiu}, {Cho}, {Clocchiatti}, {Crawford}, {Crites},
  {Desai}, {Dietrich}, {Dobbs}, {Foley}, {Forman}, {George}, {Gladders},
  {Gonzalez}, {Halverson}, {Hennig}, {Hoekstra}, {Holder}, {Holzapfel},
  {Hrubes}, {Jones}, {Keisler}, {Knox}, {Lee}, {Leitch}, {Liu}, {Lueker},
  {Luong-Van}, {Mantz}, {Marrone}, {McDonald}, {McMahon}, {Meyer}, {Mocanu},
  {Mohr}, {Murray}, {Padin}, {Pryke}, {Reichardt}, {Rest}, {Ruel}, {Ruhl},
  {Saliwanchik}, {Saro}, {Sayre}, {Schaffer}, {Schrabback}, {Shirokoff},
  {Song}, {Spieler}, {Stanford}, {Staniszewski}, {Stark}, {Story}, {Stubbs},
  {Vanderlinde}, {Vieira}, {Vikhlinin}, {Williamson}, {Zahn}, \&
  {Zenteno}}]{SPTSZ15}
{Bleem}, L.~E., {Stalder}, B., {de Haan}, T., {et~al.} 2015, \apjs, 216, 27

\bibitem[{{B{\"o}hringer} {et~al.}(2001){B{\"o}hringer}, {Schuecker}, {Guzzo},
  {Collins}, {Voges}, {Schindler}, {Neumann}, {Cruddace}, {De Grandi},
  {Chincarini}, {Edge}, {MacGillivray}, \& {Shaver}}]{Bohringer01}
{B{\"o}hringer}, H., {Schuecker}, P., {Guzzo}, L., {et~al.} 2001, \aap, 369,
  826

\bibitem[{{Bosch} {et~al.}(2018){Bosch}, {Armstrong}, {Bickerton}, {Furusawa},
  {Ikeda}, {Koike}, {Lupton}, {Mineo}, {Price}, {Takata}, {Tanaka}, {Yasuda},
  {AlSayyad}, {Becker}, {Coulton}, {Coupon}, {Garmilla}, {Huang}, {Krughoff},
  {Lang}, {Leauthaud}, {Lim}, {Lust}, {MacArthur}, {Mandelbaum}, {Miyatake},
  {Miyazaki}, {Murata}, {More}, {Okura}, {Owen}, {Swinbank}, {Strauss},
  {Yamada}, \& {Yamanoi}}]{Bosch18}
{Bosch}, J., {Armstrong}, R., {Bickerton}, S., {et~al.} 2018, \pasj, 70, S5

\bibitem[{{Bulbul} {et~al.}(2019){Bulbul}, {Chiu}, {Mohr}, {McDonald},
  {Benson}, {Bautz}, {Bayliss}, {Bleem}, {Brodwin}, {Bocquet}, {Capasso},
  {Dietrich}, {Forman}, {Hlavacek-Larrondo}, {Holzapfel}, {Khullar}, {Klein},
  {Kraft}, {Miller}, {Reichardt}, {Saro}, {Sharon}, {Stalder}, {Schrabback}, \&
  {Stanford}}]{Bulbul19}
{Bulbul}, E., {Chiu}, I.-N., {Mohr}, J.~J., {et~al.} 2019, \apj, 871, 50

\bibitem[{{Cavaliere} \& {Fusco-Femiano}(1976)}]{Cavaliere97}
{Cavaliere}, A., \& {Fusco-Femiano}, R. 1976, \aap, 500, 95

\bibitem[{{Chon} \& {B{\"o}hringer}(2017)}]{Chon17}
{Chon}, G., \& {B{\"o}hringer}, H. 2017, \aap, 606, L4

\bibitem[{{De Luca} \& {Molendi}(2004)}]{DeLuca04}
{De Luca}, A., \& {Molendi}, S. 2004, \aap, 419, 837

\bibitem[{{Diemer} \& {Kravtsov}(2015)}]{Diemer15}
{Diemer}, B., \& {Kravtsov}, A.~V. 2015, \apj, 799, 108

\bibitem[{{Donahue} {et~al.}(2014){Donahue}, {Voit}, {Mahdavi}, {Umetsu},
  {Ettori}, {Merten}, {Postman}, {Hoffer}, {Baldi}, {Coe}, {Czakon},
  {Bartelmann}, {Benitez}, {Bouwens}, {Bradley}, {Broadhurst}, {Ford},
  {Gastaldello}, {Grillo}, {Infante}, {Jouvel}, {Koekemoer}, {Kelson}, {Lahav},
  {Lemze}, {Medezinski}, {Melchior}, {Meneghetti}, {Molino}, {Moustakas},
  {Moustakas}, {Nonino}, {Rosati}, {Sayers}, {Seitz}, {Van der Wel}, {Zheng},
  \& {Zitrin}}]{Donahue14}
{Donahue}, M., {Voit}, G.~M., {Mahdavi}, A., {et~al.} 2014, \apj, 794, 136

\bibitem[{{Eckert} {et~al.}(2011){Eckert}, {Molendi}, \& {Paltani}}]{Eckert11}
{Eckert}, D., {Molendi}, S., \& {Paltani}, S. 2011, \aap, 526, A79

\bibitem[{{Efron}(1982)}]{Efron82}
{Efron}, B. 1982, {The Jackknife, the Bootstrap and other resampling plans}

\bibitem[{{Foster} {et~al.}(2012){Foster}, {Ji}, {Smith}, \&
  {Brickhouse}}]{Foster12}
{Foster}, A.~R., {Ji}, L., {Smith}, R.~K., \& {Brickhouse}, N.~S. 2012, \apj,
  756, 128

\bibitem[{{Furusawa} {et~al.}(2018){Furusawa}, {Koike}, {Takata}, {Okura},
  {Miyatake}, {Lupton}, {Bickerton}, {Price}, {Bosch}, {Yasuda}, {Mineo},
  {Yamada}, {Miyazaki}, {Nakata}, {Koshida}, {Komiyama}, {Utsumi},
  {Kawanomoto}, {Jeschke}, {Noumaru}, {Schubert}, {Iwata}, {Finet},
  {Fujiyoshi}, {Tajitsu}, {Terai}, \& {Lee}}]{Furusawa18}
{Furusawa}, H., {Koike}, M., {Takata}, T., {et~al.} 2018, \pasj, 70, S3

\bibitem[{{Giles} {et~al.}(2016){Giles}, {Maughan}, {Pacaud}, {Lieu}, {Clerc},
  {Pierre}, {Adami}, {Chiappetti}, {D{\'e}mocl{\'e}s}, {Ettori}, {Le
  F{\'e}vre}, {Ponman}, {Sadibekova}, {Smith}, {Willis}, \& {Ziparo}}]{Giles16}
{Giles}, P.~A., {Maughan}, B.~J., {Pacaud}, F., {et~al.} 2016, \aap, 592, A3

\bibitem[{{Giodini} {et~al.}(2013){Giodini}, {Lovisari}, {Pointecouteau},
  {Ettori}, {Reiprich}, \& {Hoekstra}}]{Giodini13}
{Giodini}, S., {Lovisari}, L., {Pointecouteau}, E., {et~al.} 2013, \ssr, 177,
  247

\bibitem[{{Hicks} {et~al.}(2008){Hicks}, {Ellingson}, {Bautz}, {Cain},
  {Gilbank}, {Gladders}, {Hoekstra}, {Yee}, \& {Garmire}}]{Hicks08}
{Hicks}, A.~K., {Ellingson}, E., {Bautz}, M., {et~al.} 2008, \apj, 680, 1022

\bibitem[{{Hicks} {et~al.}(2013){Hicks}, {Pratt}, {Donahue}, {Ellingson},
  {Gladders}, {B{\"o}hringer}, {Yee}, {Yan}, {Croston}, \& {Gilbank}}]{Hicks13}
{Hicks}, A.~K., {Pratt}, G.~W., {Donahue}, M., {et~al.} 2013, \mnras, 431, 2542

\bibitem[{{Hilton} {et~al.}(2017){Hilton}, {Hasselfield}, {Sif{\'o}n},
  {Battaglia}, {Aiola}, {Bharadwaj}, {Bond}, {Choi}, {Crichton}, {Datta},
  {Devlin}, {Dunkley}, {D{\"u}nner}, {Gallardo}, {Gralla}, {Hincks}, {Ho},
  {Hubmayr}, {Huffenberger}, {Hughes}, {Koopman}, {Kosowsky}, {Louis},
  {Madhavacheril}, {Marriage}, {Maurin}, {McMahon}, {Miyatake}, {Moodley},
  {Naess}, {Nati}, {Newburgh}, {Niemack}, {Oguri}, {Page}, {Partridge},
  {Schmitt}, {Sievers}, {Spergel}, {Staggs}, {Trac}, {van Engelen},
  {Vavagiakis}, \& {Wollack}}]{Hilton17}
{Hilton}, M., {Hasselfield}, M., {Sif{\'o}n}, C., {et~al.} 2017, ArXiv e-prints

\bibitem[{{Hoekstra} {et~al.}(2015){Hoekstra}, {Herbonnet}, {Muzzin}, {Babul},
  {Mahdavi}, {Viola}, \& {Cacciato}}]{Hoekstra15}
{Hoekstra}, H., {Herbonnet}, R., {Muzzin}, A., {et~al.} 2015, \mnras, 449, 685

\bibitem[{{Kalberla} {et~al.}(2005){Kalberla}, {Burton}, {Hartmann}, {Arnal},
  {Bajaja}, {Morras}, \& {P{\"o}ppel}}]{Kalberla05}
{Kalberla}, P.~M.~W., {Burton}, W.~B., {Hartmann}, D., {et~al.} 2005, \aap,
  440, 775

\bibitem[{{Katayama} {et~al.}(2003){Katayama}, {Hayashida}, {Takahara}, \&
  {Fujita}}]{Katayama03}
{Katayama}, H., {Hayashida}, K., {Takahara}, F., \& {Fujita}, Y. 2003, \apj,
  585, 687

\bibitem[{{Kawanomoto} {et~al.}(2018){Kawanomoto}, {Uraguchi}, {Komiyama},
  {Miyazaki}, {Furusawa}, {Finet}, {Hattori}, {Wang}, \&
  {Suzuki}}]{Kawanomoto18}
{Kawanomoto}, S., {Uraguchi}, F., {Komiyama}, Y., {et~al.} 2018, in prep.

\bibitem[{{Kelly}(2007)}]{Kelly07}
{Kelly}, B.~C. 2007, \apj, 665, 1489

\bibitem[{{Komiyama} {et~al.}(2018){Komiyama}, {Obuchi}, {Nakaya}, {Kamata},
  {Kawanomoto}, {Utsumi}, {Miyazaki}, {Uraguchi}, {Furusawa}, {Morokuma},
  {Uchida}, {Miyatake}, {Mineo}, {Fujimori}, {Aihara}, {Karoji}, {Gunn}, \&
  {Wang}}]{Komiyama18}
{Komiyama}, Y., {Obuchi}, Y., {Nakaya}, H., {et~al.} 2018, \pasj, 70, S2

\bibitem[{{Lodders} \& {Palme}(2009)}]{Lodders09}
{Lodders}, K., \& {Palme}, H. 2009, Meteoritics and Planetary Science
  Supplement, 72, 5154

\bibitem[{{Lovisari} {et~al.}(2017){Lovisari}, {Forman}, {Jones}, {Ettori},
  {Andrade-Santos}, {Arnaud}, {D{\'e}mocl{\`e}s}, {Pratt}, {Randall}, \&
  {Kraft}}]{Lovisari17}
{Lovisari}, L., {Forman}, W.~R., {Jones}, C., {et~al.} 2017, \apj, 846, 51

\bibitem[{{Mahdavi} {et~al.}(2013){Mahdavi}, {Hoekstra}, {Babul}, {Bildfell},
  {Jeltema}, \& {Henry}}]{Mahdavi13}
{Mahdavi}, A., {Hoekstra}, H., {Babul}, A., {et~al.} 2013, \apj, 767, 116

\bibitem[{{Mann} \& {Ebeling}(2012)}]{Mann12}
{Mann}, A.~W., \& {Ebeling}, H. 2012, \mnras, 420, 2120

\bibitem[{{Mantz} {et~al.}(2016){Mantz}, {Allen}, {Morris}, {von der Linden},
  {Applegate}, {Kelly}, {Burke}, {Donovan}, \& {Ebeling}}]{Mantz16}
{Mantz}, A.~B., {Allen}, S.~W., {Morris}, R.~G., {et~al.} 2016, \mnras, 463,
  3582

\bibitem[{{Martino} {et~al.}(2014){Martino}, {Mazzotta}, {Bourdin}, {Smith},
  {Bartalucci}, {Marrone}, {Finoguenov}, \& {Okabe}}]{Martino14}
{Martino}, R., {Mazzotta}, P., {Bourdin}, H., {et~al.} 2014, \mnras, 443, 2342

\bibitem[{{Maughan} {et~al.}(2012){Maughan}, {Giles}, {Randall}, {Jones}, \&
  {Forman}}]{Maughan12}
{Maughan}, B.~J., {Giles}, P.~A., {Randall}, S.~W., {Jones}, C., \& {Forman},
  W.~R. 2012, \mnras, 421, 1583

\bibitem[{{McDonald} {et~al.}(2017){McDonald}, {Allen}, {Bayliss}, {Benson},
  {Bleem}, {Brodwin}, {Bulbul}, {Carlstrom}, {Forman}, {Hlavacek-Larrondo},
  {Garmire}, {Gaspari}, {Gladders}, {Mantz}, \& {Murray}}]{McDonald17}
{McDonald}, M., {Allen}, S.~W., {Bayliss}, M., {et~al.} 2017, \apj, 843, 28

\bibitem[{{Merloni} {et~al.}(2012){Merloni}, {Predehl}, {Becker},
  {B{\"o}hringer}, {Boller}, {Brunner}, {Brusa}, {Dennerl}, {Freyberg},
  {Friedrich}, {Georgakakis}, {Haberl}, {Hasinger}, {Meidinger}, {Mohr},
  {Nandra}, {Rau}, {Reiprich}, {Robrade}, {Salvato}, {Santangelo}, {Sasaki},
  {Schwope}, {Wilms}, \& {German eROSITA Consortium}}]{Merloni12}
{Merloni}, A., {Predehl}, P., {Becker}, W., {et~al.} 2012, ArXiv e-prints

\bibitem[{{Miyaoka} {et~al.}(2018){Miyaoka}, {Okabe}, {Kitaguchi}, {Oguri},
  {Fukazawa}, {Mandelbaum}, {Medezinski}, {Babazaki}, {Nishizawa}, {Hamana},
  {Lin}, {Akamatsu}, {Chiu}, {Fujita}, {Ichinohe}, {Komiyama}, {Sasaki},
  {Takizawa}, {Ueda}, {Umetsu}, {Coupon}, {Hikage}, {Hoshino}, {Leauthaud},
  {Matsushita}, {Mitsuishi}, {Miyatake}, {Miyazaki}, {More}, {Nakazawa}, {Ota},
  {Sato}, {Spergel}, {Tamura}, {Tanaka}, {Tanaka}, \& {Utsumi}}]{Miyaoka18}
{Miyaoka}, K., {Okabe}, N., {Kitaguchi}, T., {et~al.} 2018, \pasj, 70, S22

\bibitem[{{Miyazaki} {et~al.}(2012){Miyazaki}, {Komiyama}, {Nakaya}, {Kamata},
  {Doi}, {Hamana}, {Karoji}, {Furusawa}, {Kawanomoto}, {Morokuma}, {Ishizuka},
  {Nariai}, {Tanaka}, {Uraguchi}, {Utsumi}, {Obuchi}, {Okura}, {Oguri},
  {Takata}, {Tomono}, {Kurakami}, {Namikawa}, {Usuda}, {Yamanoi}, {Terai},
  {Uekiyo}, {Yamada}, {Koike}, {Aihara}, {Fujimori}, {Mineo}, {Miyatake},
  {Yasuda}, {Nishizawa}, {Saito}, {Tanaka}, {Uchida}, {Katayama}, {Wang},
  {Chen}, {Lupton}, {Loomis}, {Bickerton}, {Price}, {Gunn}, {Suzuki},
  {Miyazaki}, {Muramatsu}, {Yamamoto}, {Endo}, {Ezaki}, {Itoh}, {Miwa},
  {Yokota}, {Matsuda}, {Ebinuma}, \& {Takeshi}}]{Miyazaki12}
{Miyazaki}, S., {Komiyama}, Y., {Nakaya}, H., {et~al.} 2012, in \procspie, Vol.
  8446, Ground-based and Airborne Instrumentation for Astronomy IV, 84460Z

\bibitem[{{Miyazaki} {et~al.}(2015){Miyazaki}, {Oguri}, {Hamana}, {Tanaka},
  {Miller}, {Utsumi}, {Komiyama}, {Furusawa}, {Sakurai}, {Kawanomoto},
  {Nakata}, {Uraguchi}, {Koike}, {Tomono}, {Lupton}, {Gunn}, {Karoji},
  {Aihara}, {Murayama}, \& {Takada}}]{Miyazaki15}
{Miyazaki}, S., {Oguri}, M., {Hamana}, T., {et~al.} 2015, \apj, 807, 22

\bibitem[{{Miyazaki} {et~al.}(2018){Miyazaki}, {Komiyama}, {Kawanomoto}, {Doi},
  {Furusawa}, {Hamana}, {Hayashi}, {Ikeda}, {Kamata}, {Karoji}, {Koike},
  {Kurakami}, {Miyama}, {Morokuma}, {Nakata}, {Namikawa}, {Nakaya}, {Nariai},
  {Obuchi}, {Oishi}, {Okada}, {Okura}, {Tait}, {Takata}, {Tanaka}, {Tanaka},
  {Terai}, {Tomono}, {Uraguchi}, {Usuda}, {Utsumi}, {Yamada}, {Yamanoi},
  {Aihara}, {Fujimori}, {Mineo}, {Miyatake}, {Oguri}, {Uchida}, {Tanaka},
  {Yasuda}, {Takada}, {Murayama}, {Nishizawa}, {Sugiyama}, {Chiba}, {Futamase},
  {Wang}, {Chen}, {Ho}, {Liaw}, {Chiu}, {Ho}, {Lai}, {Lee}, {Jeng}, {Iwamura},
  {Armstrong}, {Bickerton}, {Bosch}, {Gunn}, {Lupton}, {Loomis}, {Price},
  {Smith}, {Strauss}, {Turner}, {Suzuki}, {Miyazaki}, {Muramatsu}, {Yamamoto},
  {Endo}, {Ezaki}, {Ito}, {Kawaguchi}, {Sofuku}, {Taniike}, {Akutsu}, {Dojo},
  {Kasumi}, {Matsuda}, {Imoto}, {Miwa}, {Suzuki}, {Takeshi}, \&
  {Yokota}}]{Miyazaki18}
{Miyazaki}, S., {Komiyama}, Y., {Kawanomoto}, S., {et~al.} 2018, \pasj, 70, S1

\bibitem[{{Oguri}(2014)}]{Oguri14b}
{Oguri}, M. 2014, \mnras, 444, 147

\bibitem[{{Oguri} {et~al.}(2018){Oguri}, {Lin}, {Lin}, {Nishizawa}, {More},
  {More}, {Hsieh}, {Medezinski}, {Miyatake}, {Jian}, {Lin}, {Takada}, {Okabe},
  {Speagle}, {Coupon}, {Leauthaud}, {Lupton}, {Miyazaki}, {Price}, {Tanaka},
  {Chiu}, {Komiyama}, {Okura}, {Tanaka}, \& {Usuda}}]{Oguri18}
{Oguri}, M., {Lin}, Y.-T., {Lin}, S.-C., {et~al.} 2018, \pasj, 70, S20

\bibitem[{{Okabe} {et~al.}(2014){Okabe}, {Umetsu}, {Tamura}, {Fujita},
  {Takizawa}, {Zhang}, {Matsushita}, {Hamana}, {Fukazawa}, {Futamase},
  {Kawaharada}, {Miyazaki}, {Mochizuki}, {Nakazawa}, {Ohashi}, {Ota}, {Sasaki},
  {Sato}, \& {Tam}}]{Okabe14b}
{Okabe}, N., {Umetsu}, K., {Tamura}, T., {et~al.} 2014, \pasj, 66, 99

\bibitem[{{Okabe} {et~al.}(2018){Okabe}, {Oguri}, {Akamatsu}, {Hamabata},
  {Nishizawa}, {Medezinski}, {Koyama}, {Hayashi}, {Okabe}, {Ueda}, {Mitsuishi},
  \& {Ota}}]{Okabe18}
{Okabe}, N., {Oguri}, M., {Akamatsu}, H., {et~al.} 2018, arXiv e-prints

\bibitem[{{Ota} {et~al.}(2006){Ota}, {Kitayama}, {Masai}, \& {Mitsuda}}]{Ota06}
{Ota}, N., {Kitayama}, T., {Masai}, K., \& {Mitsuda}, K. 2006, \apj, 640, 673

\bibitem[{{Ota} \& {Mitsuda}(2004)}]{Ota04}
{Ota}, N., \& {Mitsuda}, K. 2004, \aap, 428, 757

\bibitem[{{Pierre} {et~al.}(2004){Pierre}, {Valtchanov}, {Altieri}, {Andreon},
  {Bolzonella}, {Bremer}, {Disseau}, {Dos Santos}, {Gandhi}, {Jean}, {Pacaud},
  {Read}, {Refregier}, {Willis}, {Adami}, {Alloin}, {Birkinshaw}, {Chiappetti},
  {Cohen}, {Detal}, {Duc}, {Gosset}, {Hjorth}, {Jones}, {Le F{\`e}vre},
  {Lonsdale}, {Maccagni}, {Mazure}, {McBreen}, {McCracken}, {Mellier},
  {Ponman}, {Quintana}, {Rottgering}, {Smette}, {Surdej}, {Starck}, {Vigroux},
  \& {White}}]{Pierre04}
{Pierre}, M., {Valtchanov}, I., {Altieri}, B., {et~al.} 2004, \jcap, 9, 011

\bibitem[{{Pierre} {et~al.}(2016{\natexlab{a}}){Pierre}, {Pacaud}, {Adami},
  {Alis}, {Altieri}, {Baran}, {Benoist}, {Birkinshaw}, {Bongiorno}, {Bremer},
  {Brusa}, {Butler}, {Ciliegi}, {Chiappetti}, {Clerc}, {Corasaniti}, {Coupon},
  {De Breuck}, {Democles}, {Desai}, {Delhaize}, {Devriendt}, {Dubois},
  {Eckert}, {Elyiv}, {Ettori}, {Evrard}, {Faccioli}, {Farahi}, {Ferrari},
  {Finet}, {Fotopoulou}, {Fourmanoit}, {Gandhi}, {Gastaldello}, {Gastaud},
  {Georgantopoulos}, {Giles}, {Guennou}, {Guglielmo}, {Horellou}, {Husband},
  {Huynh}, {Iovino}, {Kilbinger}, {Koulouridis}, {Lavoie}, {Le Brun}, {Le
  Fevre}, {Lidman}, {Lieu}, {Lin}, {Mantz}, {Maughan}, {Maurogordato},
  {McCarthy}, {McGee}, {Melin}, {Melnyk}, {Menanteau}, {Novak}, {Paltani},
  {Plionis}, {Poggianti}, {Pomarede}, {Pompei}, {Ponman}, {Ramos-Ceja},
  {Ranalli}, {Rapetti}, {Raychaudury}, {Reiprich}, {Rottgering}, {Rozo},
  {Rykoff}, {Sadibekova}, {Santos}, {Sauvageot}, {Schimd}, {Sereno}, {Smith},
  {Smol{\v c}i{\'c}}, {Snowden}, {Spergel}, {Stanford}, {Surdej}, {Valageas},
  {Valotti}, {Valtchanov}, {Vignali}, {Willis}, \& {Ziparo}}]{XXL16}
{Pierre}, M., {Pacaud}, F., {Adami}, C., {et~al.} 2016{\natexlab{a}}, \aap,
  592, A1

\bibitem[{{Pierre} {et~al.}(2016{\natexlab{b}}){Pierre}, {Pacaud}, {Adami},
  {Alis}, {Altieri}, {Baran}, {Benoist}, {Birkinshaw}, {Bongiorno}, {Bremer},
  {Brusa}, {Butler}, {Ciliegi}, {Chiappetti}, {Clerc}, {Corasaniti}, {Coupon},
  {De Breuck}, {Democles}, {Desai}, {Delhaize}, {Devriendt}, {Dubois},
  {Eckert}, {Elyiv}, {Ettori}, {Evrard}, {Faccioli}, {Farahi}, {Ferrari},
  {Finet}, {Fotopoulou}, {Fourmanoit}, {Gandhi}, {Gastaldello}, {Gastaud},
  {Georgantopoulos}, {Giles}, {Guennou}, {Guglielmo}, {Horellou}, {Husband},
  {Huynh}, {Iovino}, {Kilbinger}, {Koulouridis}, {Lavoie}, {Le Brun}, {Le
  Fevre}, {Lidman}, {Lieu}, {Lin}, {Mantz}, {Maughan}, {Maurogordato},
  {McCarthy}, {McGee}, {Melin}, {Melnyk}, {Menanteau}, {Novak}, {Paltani},
  {Plionis}, {Poggianti}, {Pomarede}, {Pompei}, {Ponman}, {Ramos-Ceja},
  {Ranalli}, {Rapetti}, {Raychaudury}, {Reiprich}, {Rottgering}, {Rozo},
  {Rykoff}, {Sadibekova}, {Santos}, {Sauvageot}, {Schimd}, {Sereno}, {Smith},
  {Smol{\v c}i{\'c}}, {Snowden}, {Spergel}, {Stanford}, {Surdej}, {Valageas},
  {Valotti}, {Valtchanov}, {Vignali}, {Willis}, \& {Ziparo}}]{Pierre16}
---. 2016{\natexlab{b}}, \aap, 592, A1

\bibitem[{{Reichert} {et~al.}(2011){Reichert}, {B{\"o}hringer}, {Fassbender},
  \& {M{\"u}hlegger}}]{Reichert11}
{Reichert}, A., {B{\"o}hringer}, H., {Fassbender}, R., \& {M{\"u}hlegger}, M.
  2011, \aap, 535, A4

\bibitem[{{Rosen} {et~al.}(2016){Rosen}, {Webb}, {Watson}, {Ballet}, {Barret},
  {Braito}, {Carrera}, {Ceballos}, {Coriat}, {Della Ceca}, {Denkinson},
  {Esquej}, {Farrell}, {Freyberg}, {Gris{\'e}}, {Guillout}, {Heil},
  {Koliopanos}, {Law-Green}, {Lamer}, {Lin}, {Martino}, {Michel}, {Motch},
  {Nebot Gomez-Moran}, {Page}, {Page}, {Page}, {Pakull}, {Pye}, {Read},
  {Rodriguez}, {Sakano}, {Saxton}, {Schwope}, {Scott}, {Sturm}, {Traulsen},
  {Yershov}, \& {Zolotukhin}}]{Rosen16}
{Rosen}, S.~R., {Webb}, N.~A., {Watson}, M.~G., {et~al.} 2016, \aap, 590, A1

\bibitem[{{Rossetti} {et~al.}(2017){Rossetti}, {Gastaldello}, {Eckert}, {Della
  Torre}, {Pantiri}, {Cazzoletti}, \& {Molendi}}]{Rossetti17}
{Rossetti}, M., {Gastaldello}, F., {Eckert}, D., {et~al.} 2017, \mnras, 468,
  1917

\bibitem[{{Rossetti} {et~al.}(2016){Rossetti}, {Gastaldello}, {Ferioli},
  {Bersanelli}, {De Grandi}, {Eckert}, {Ghizzardi}, {Maino}, \&
  {Molendi}}]{Rossetti16}
{Rossetti}, M., {Gastaldello}, F., {Ferioli}, G., {et~al.} 2016, \mnras, 457,
  4515

\bibitem[{{Rozo} \& {Rykoff}(2014)}]{Rozo14}
{Rozo}, E., \& {Rykoff}, E.~S. 2014, \apj, 783, 80

\bibitem[{{Rykoff} {et~al.}(2014){Rykoff}, {Rozo}, {Busha}, {Cunha},
  {Finoguenov}, {Evrard}, {Hao}, {Koester}, {Leauthaud}, {Nord}, {Pierre},
  {Reddick}, {Sadibekova}, {Sheldon}, \& {Wechsler}}]{Rykoff14}
{Rykoff}, E.~S., {Rozo}, E., {Busha}, M.~T., {et~al.} 2014, \apj, 785, 104

\bibitem[{{Rykoff} {et~al.}(2016){Rykoff}, {Rozo}, {Hollowood},
  {Bermeo-Hernandez}, {Jeltema}, {Mayers}, {Romer}, {Rooney}, {Saro}, {Vergara
  Cervantes}, {Wechsler}, {Wilcox}, {Abbott}, {Abdalla}, {Allam}, {Annis},
  {Benoit-L{\'e}vy}, {Bernstein}, {Bertin}, {Brooks}, {Burke}, {Capozzi},
  {Carnero Rosell}, {Carrasco Kind}, {Castander}, {Childress}, {Collins},
  {Cunha}, {D'Andrea}, {da Costa}, {Davis}, {Desai}, {Diehl}, {Dietrich},
  {Doel}, {Evrard}, {Finley}, {Flaugher}, {Fosalba}, {Frieman}, {Glazebrook},
  {Goldstein}, {Gruen}, {Gruendl}, {Gutierrez}, {Hilton}, {Honscheid}, {Hoyle},
  {James}, {Kay}, {Kuehn}, {Kuropatkin}, {Lahav}, {Lewis}, {Lidman}, {Lima},
  {Maia}, {Mann}, {Marshall}, {Martini}, {Melchior}, {Miller}, {Miquel},
  {Mohr}, {Nichol}, {Nord}, {Ogando}, {Plazas}, {Reil}, {Sahl{\'e}n},
  {Sanchez}, {Santiago}, {Scarpine}, {Schubnell}, {Sevilla-Noarbe}, {Smith},
  {Soares-Santos}, {Sobreira}, {Stott}, {Suchyta}, {Swanson}, {Tarle},
  {Thomas}, {Tucker}, {Uddin}, {Viana}, {Vikram}, {Walker}, {Zhang}, \& {DES
  Collaboration}}]{Rykoff16}
{Rykoff}, E.~S., {Rozo}, E., {Hollowood}, D., {et~al.} 2016, \apjs, 224, 1

\bibitem[{{Sanders} {et~al.}(2018){Sanders}, {Fabian}, {Russell}, \&
  {Walker}}]{Sanders18}
{Sanders}, J.~S., {Fabian}, A.~C., {Russell}, H.~R., \& {Walker}, S.~A. 2018,
  \mnras, 474, 1065

\bibitem[{{Sanderson} {et~al.}(2009){Sanderson}, {Edge}, \&
  {Smith}}]{Sanders09}
{Sanderson}, A.~J.~R., {Edge}, A.~C., \& {Smith}, G.~P. 2009, \mnras, 398, 1698

\bibitem[{{Smith} {et~al.}(2016){Smith}, {Mazzotta}, {Okabe}, {Ziparo},
  {Mulroy}, {Babul}, {Finoguenov}, {McCarthy}, {Lieu}, {Bah{\'e}}, {Bourdin},
  {Evrard}, {Futamase}, {Haines}, {Jauzac}, {Marrone}, {Martino}, {May},
  {Taylor}, \& {Umetsu}}]{Smith16}
{Smith}, G.~P., {Mazzotta}, P., {Okabe}, N., {et~al.} 2016, \mnras, 456, L74

\bibitem[{{Smith} {et~al.}(2001){Smith}, {Brickhouse}, {Liedahl}, \&
  {Raymond}}]{Smith01}
{Smith}, R.~K., {Brickhouse}, N.~S., {Liedahl}, D.~A., \& {Raymond}, J.~C.
  2001, \apjl, 556, L91

\bibitem[{{Snowden} {et~al.}(2008){Snowden}, {Mushotzky}, {Kuntz}, \&
  {Davis}}]{Snowden08}
{Snowden}, S.~L., {Mushotzky}, R.~F., {Kuntz}, K.~D., \& {Davis}, D.~S. 2008,
  \aap, 478, 615

\bibitem[{{Snowden} {et~al.}(1997){Snowden}, {Egger}, {Freyberg}, {McCammon},
  {Plucinsky}, {Sanders}, {Schmitt}, {Tr{\"u}mper}, \& {Voges}}]{Snowden97}
{Snowden}, S.~L., {Egger}, R., {Freyberg}, M.~J., {et~al.} 1997, \apj, 485, 125

\bibitem[{{Sun} {et~al.}(2009){Sun}, {Voit}, {Donahue}, {Jones}, {Forman}, \&
  {Vikhlinin}}]{Sun09}
{Sun}, M., {Voit}, G.~M., {Donahue}, M., {et~al.} 2009, \apj, 693, 1142

\bibitem[{{Takey} {et~al.}(2011){Takey}, {Schwope}, \& {Lamer}}]{Takey11}
{Takey}, A., {Schwope}, A., \& {Lamer}, G. 2011, \aap, 534, A120

\bibitem[{{Takey} {et~al.}(2013){Takey}, {Schwope}, \& {Lamer}}]{Takey13}
---. 2013, \aap, 558, A75

\bibitem[{{Tanaka} {et~al.}(2018){Tanaka}, {Coupon}, {Hsieh}, {Mineo},
  {Nishizawa}, {Speagle}, {Furusawa}, {Miyazaki}, \& {Murayama}}]{HSCPhotoz18}
{Tanaka}, M., {Coupon}, J., {Hsieh}, B.-C., {et~al.} 2018, \pasj, 70, S9

\bibitem[{{Vikhlinin} {et~al.}(2006){Vikhlinin}, {Kravtsov}, {Forman}, {Jones},
  {Markevitch}, {Murray}, \& {Van Speybroeck}}]{Vikhlinin06}
{Vikhlinin}, A., {Kravtsov}, A., {Forman}, W., {et~al.} 2006, \apj, 640, 691

\bibitem[{{von der Linden} {et~al.}(2014){von der Linden}, {Mantz}, {Allen},
  {Applegate}, {Kelly}, {Morris}, {Wright}, {Allen}, {Burchat}, {Burke},
  {Donovan}, \& {Ebeling}}]{vonderLinden14}
{von der Linden}, A., {Mantz}, A., {Allen}, S.~W., {et~al.} 2014, \mnras, 443,
  1973

\bibitem[{{Zhang} {et~al.}(2008){Zhang}, {Finoguenov}, {B{\"o}hringer},
  {Kneib}, {Smith}, {Kneissl}, {Okabe}, \& {Dahle}}]{Zhang08}
{Zhang}, Y.-Y., {Finoguenov}, A., {B{\"o}hringer}, H., {et~al.} 2008, \aap,
  482, 451

\end{thebibliography}

\end{document}